\documentclass[aps,prb,twocolumn,groupedaddress,showpacs,floatfix,superscriptaddress,longbibliography]{revtex4-2}
\usepackage[plainpages=false,pdfpagelabels,colorlinks=true,linkcolor=red,urlcolor=blue,citecolor=blue,pdftitle={Title},pdfauthor={Authors},pdfdisplaydoctitle=true,pdfduplex=DuplexFlipLongEdge]{hyperref}
\usepackage{epsfig}
\usepackage{amsmath,amssymb}
\usepackage{physics}
\usepackage{graphicx}
\usepackage[dvipsnames,usenames]{color}
\usepackage[normalem]{ulem}
\usepackage{soul} 
\usepackage{float}
\usepackage{diagbox}
\usepackage{orcidlink}
\usepackage[version=4]{mhchem}

\def\s{\sigma}
\def\ave#1{\langle #1\rangle}

\newcommand{\iv}{\mathbf{i}}
\newcommand{\jv}{\mathbf{j}}
\newcommand{\qv}{\mathbf{q}}

\newcommand{\kv}{\mathbf{k}} 

\begin{document}

\title {Spin-orbit coupled periodic Anderson model: Kondo-Dirac semimetal and orbital-selective antiferromagnetic semimetal}

\author{Sebasti\~ao dos Anjos Sousa-Júnior \orcidlink{0000-0002-4266-3780}}
\affiliation{Department of Physics, University of Houston, Houston, Texas 77204, USA}
\author{Julián Faúndez \orcidlink{0000-0002-6909-0417}}
\affiliation{Departamento de Física y Astronomía, Universidad Andrés Bello, Santiago 837-0136, Chile}
\author{Rubem Mondaini \orcidlink{0000-0001-8005-2297}}
\affiliation{Department of Physics, University of Houston, Houston, Texas 77204, USA}
\affiliation{Texas Center for Superconductivity, University of Houston, Houston, Texas 77204, USA}

\begin{abstract}

We investigate the periodic Anderson model composed of an itinerant $c$-band and a strongly localized $f$-band, featuring on-site electron-electron interactions in the $f$-orbitals. The two bands interact via a hybridization term with spin-orbit coupling, which enables spin-flip processes. In the non-interacting limit, these profoundly alter the electronic structure, leading to the emergence of flat bands, van Hove singularities, and, most notably, Dirac cones within a single Kondo-Dirac semimetal order. The strongly interacting regime is explored via the determinant quantum Monte Carlo method, in the absence of the sign problem, where we unveil a complete ground-state phase diagram revealing two distinct phases, the Kondo-Dirac semimetal phase and a novel antiferromagnetic semimetal phase. Their characterization by the spectral functions establishes an orbital-selective Mott transition in the antiferromagnetic semimetal phase, marked by the opening of a gap exclusively in the $f$-orbital while Dirac cones persist in the $c$-orbital. Conversely, in the Kondo-Dirac semimetal phase, both $c$- and $f$-orbitals sustain robust Dirac cones. We establish that spin-orbit coupling in the hybridization term gives rise to Dirac cones, which, combined with additional symmetry-breaking conditions, can generate novel topological states. 
   
\end{abstract}

\maketitle


\section{Introduction}

Heavy-fermion systems constitute a class of strongly correlated electron materials in which hybridization between itinerant conduction electrons and localized $f$-electron orbitals gives rise to heavy quasiparticles with effective masses that can exceed hundreds of times the bare electron mass~\cite{Coleman2007}. These materials, commonly based on rare-earth or actinide elements such as Ce, Yb, or U, exhibit a rich spectrum of emergent quantum phenomena at low temperatures, such as superconductivity under applied pressure~\cite{Kimura2005}, unconventional superconducting pairing mechanisms~\cite{Pfleiderer2009}, complex magnetic ordering~\cite{Kuramoto2009}, non-Fermi-liquid behavior near quantum critical points~\cite{Lohneysen2007}, and exotic phases such as the `hidden order' observed in \ce{URu2Si2}~\cite{Mydosh2011}. These diverse ground states arise from a delicate competition between Kondo screening~\cite{Kondo1964}, which favors itinerant $c$-$f$ singlet formation, and the Ruderman-Kittel-Kasuya-Yosida (RKKY) interaction~\cite{Ruderman1954, Kasuya1956, Yosida1957}, mediating long-range magnetic ordering via conduction electron spin polarization~\cite{Coleman2007}. 

In recent years, growing attention has focused on the decisive role of spin-orbit coupling (SOC) in heavy-fermion systems~\cite{Michishita2019}. Owing to the strong atomic SOC of $f$-electron orbitals, its inclusion is indispensable for accurately describing their low-energy electronic structure. SOC not only changes the band structure but also plays a key role in shaping magnetic anisotropies, lifting orbital degeneracies, and enabling exotic topological phenomena. Experimental investigations on compounds such as \ce{Ce3Bi4Pt3}~\cite{Dzsaber2017}, \ce{SmB6}~\cite{Xu2013, Ohtsubo2022}, \ce{CeSiI}~\cite{Fumega2024}, and \ce{Yb}-based compounds~\cite{Generalov2018} have demonstrated that SOC crucially determines their low-energy spectra. The interplay between SOC and electronic correlations can result in novel quantum phases, such as Kondo semimetals, topological Kondo insulators, or even anisotropic magnetic states~\cite{Generalov2018, Mende2022, Tarasov2022}. In particular, SOC lifts orbital degeneracies and introduces spin-momentum locking, thereby modifying the nature of the hybridization gap and influencing the diverse states of matter. Additionally, the strong interactions and SOC can also drive heavy-fermion systems to Weyl-Kondo semimetal phases~\cite{Dzsaber2017, Lai2018, Grefe2020, Samir2021}. More recently, the identification of topologically nontrivial phases in certain heavy-fermion compounds has further enriched the field, spawning new interest in the study of correlated topological matter~\cite{Neupane2013, Weng2014, Dzero2016}.

\begin{figure*}[t!]
    \centering
    \includegraphics[width=1.0\linewidth]{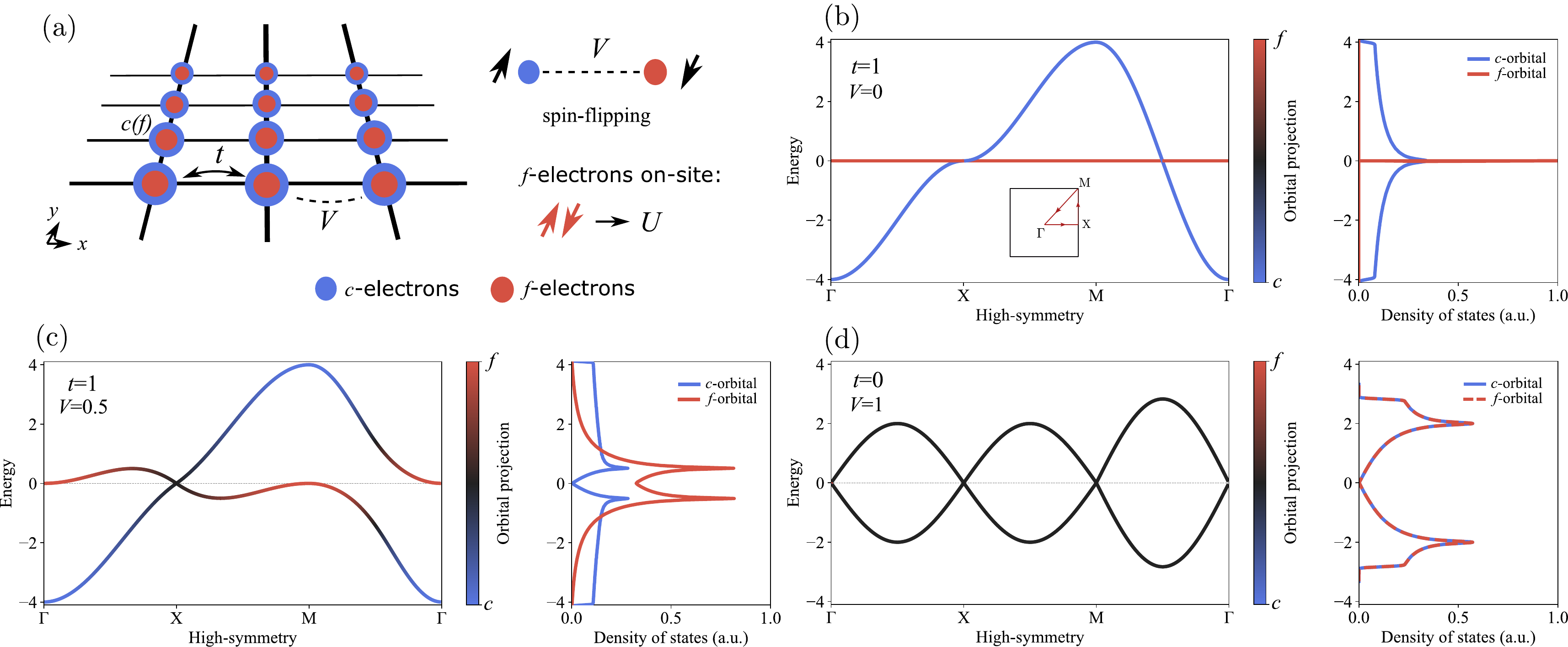}
    \caption{(a) Schematic representation of the Periodic Anderson Model (PAM) on a square lattice, where localized $f$-orbitals at each site hybridize with dispersive $c$- electrons through a hybridization $V$ that includes spin–orbit coupling (SOC). Panels (b)–(d) display the tight-binding approximation for the band structure, along a high symmetry path in the Brillouin zone [inset in (b)], and density of states DOS for three regimes: (b) $V=0$, $t=1$; (c) intermediate SOC with $V=0.5$, $t=1$; and (d) pure SOC limit with $V=1$, $t=0$, at zero temperature. Both are resolved by the contributions of the two types of orbitals.
    }
    \label{fig1}
\end{figure*}

From a theoretical perspective, the periodic Anderson model (PAM), the lattice generalization of the single-impurity Anderson model, plays a central role in the study of systems comprising two electronic bands, one itinerant and one strongly localized. It is particularly relevant in the context of heavy-fermion materials, where hybridization between conduction electrons and localized $f$-electrons gives rise to a wide range of many-body phenomena~\cite{Vidhyadhiraja2004, Hagymasi2011, Costa2018, Costa2019, Zhang2019, Oliveira2023, Gleis2024, Hagymasi2025}. Extensions of the PAM have incorporated SOC, motivated by its importance in materials with strong relativistic effects and complex orbital structures. The inclusion of SOC has revealed intrinsic spin and orbital Hall effects in heavy-fermion systems~\cite{Tanaka2010}, as well as anisotropic, momentum-dependent exchange interactions that stabilize unconventional magnetic textures~\cite{Yambe2022}. Additionally, it has led to modifications in magnetic susceptibility due to spin-orbit entanglement~\cite{Harima1987}. Moreover, variants of the Anderson lattice model have been extensively studied in the context of emergent topological phases in these systems~\cite{Dzero2010, Dzero2012, Legner2014, Lu2019}.

In the context of numerical simulations, the PAM has been previously studied using quantum Monte Carlo (QMC) methods~\cite{Costa2018, Costa2019, Zhang2019}. Furthermore, the interplay between SOC and strong electronic interactions has been extensively explored in Rashba systems, both for attractive \cite{Tang2014, Rosenberg2017} and repulsive \cite{Kim2020, Wan2022, SousaJunior2025} interactions, as well as in models incorporating electron-phonon coupling \cite{Faundez2024}. However, the effects of SOC in the PAM have, to date, been investigated primarily at the mean-field level in three-dimensional systems \cite{Legner2014}, or using QMC in one- and two-dimensional variants of the model \cite{Luo2021}. 

Motivated by the experimental and theoretical developments discussed above, we perform a numerical investigation of the PAM on a square lattice using the finite-temperature determinant quantum Monte Carlo (DQMC) method~\cite{Blankenbecler1981}. Our analysis focuses on the regime in which SOC is incorporated into the hybridization between the itinerant conduction band and the strongly interacting localized $f$-orbitals. This article is organized as follows. In Sec.~\ref{sec:Method}, we introduce the theoretical framework and model Hamiltonian, along with a description of the DQMC methodology employed in our simulations. Section~\ref{sec:Results} presents our main numerical results, including the analysis of spectral properties of each emergent phase and the determination of the ground-state phase boundaries. Finally, in Sec.~\ref{sec:Conclusions}, we summarize our findings and discuss their broader implications.


\section{Model and Method}
\label{sec:Method}

The Hamiltonian of the PAM with SOC is given as,
\begin{align}
    \mathcal{H}= &-t\sum_{\langle \iv ,\jv\rangle}\left(c^\dagger_\iv c_\jv^{\phantom{\dagger}}+{\rm H.c.}\right) + U\sum_\iv \,\left(n^f_{\iv\uparrow}-\dfrac{1}{2}\right)\left(\,n^f_{\iv\downarrow}-\dfrac{1}{2}\right), \nonumber \\
    & + \sum_{\langle \iv,\jv \rangle} \left[i V c_\iv^\dagger\sigma_\alpha f^{}_\jv + i V f_\iv^\dagger\sigma_\alpha c_\jv^{\phantom{\dagger}}\!+\!{\rm H.c.}\right],
\label{eq:Hamiltonian}
\end{align}
where $c_\iv$, $f_\iv$ ($c_\iv^\dagger$, $f_\iv^\dagger$) are the annihilation (creation) operators for the conduction $c$- and local $f$-electrons, respectively. In the same way, these are defined as spinors of the form $c_\iv^\dagger\equiv(c_{\iv\uparrow}^\dagger,c_{\iv\downarrow}^\dagger)$ and $f_\iv^\dagger\equiv(f_{\iv\uparrow}^\dagger,f_{\iv\downarrow}^\dagger)$, encompassing the spin-degrees of freedom $\sigma$. Here, $t$ is the nearest-neighbor hopping amplitude for the itinerant $c$-electrons; $U$ the on-site electron-electron interaction between $f$-electrons, $n^f_{\iv\sigma} = f_{\iv\sigma}^\dagger f_{\iv\sigma}^{\phantom{\dagger}}$ is the $f$-electron number operator for spin $\sigma$ at site $\iv$, and finally $V$ is the hybridization amplitude between $c$- and $f$-electrons. The matrices $\s_\alpha$ are the Pauli matrices associated with the spin component $\alpha={x,y}$ depending on the direction of the hopping between sites $\iv$ and $\jv$. Here, the sums are over sites of a square lattice with periodic boundary conditions, and $\ave{\iv,\jv}$ denotes nearest-neighbor sites. Schematically, the relevant terms of the model are represented in Fig.~\ref{fig1}(a).

We analyze the ground state physics of Eq.\,\eqref{eq:Hamiltonian} by employing a finite-temperature DQMC method \cite{Blankenbecler1981,Hirsch1985,rrds2003,Assaad2008,Gubernatis16}. In this approach, noncommuting terms of the Hamiltonian in the partition function are decoupled through the Trotter-Suzuki decomposition. This leads to the introduction of an imaginary-time axis through the discretization of the inverse temperature, \(\beta = L_\tau \Delta \tau\), where $L_\tau$ and $\Delta \tau$ represent the number of time slices and the discretization step, respectively. Notice that the Suzuki-Trotter decomposition introduces an error proportional to \((\Delta\tau)^2\); here, we define $t \Delta \tau =0.1$, which is sufficiently small to ensure that systematic errors are comparable to the statistical ones. To recast the quartic interaction term in quadratic form, we apply a Hubbard-Stratonovich (HS) \,\cite{Hirsch1983} transformation. This allows one to trace over fermionic degrees of freedom in the partition function, where the remaining degrees of freedom on the auxiliary fields are then evaluated via importance sampling, with a determinant serving as the configuration weight. We have that for any combination of ${U,V,t}$ the system free from the sign problem at \textit{half-filling}\,\cite{Hirsch1985,White1988,Loh1990,Mondaini2022,Mondaini2023}, owing to a combination of time-reversal and particle-hole transformations --- more on this in Appendix\,\ref{App:sign_prob}.  

We examine the emergence of magnetic long-range order by calculating the spin-spin correlations, $\langle {\bf S}^{\gamma}_{\mathbf{i}} \cdot {\bf S}^{\gamma}_{\mathbf{j}} \rangle$, for a given orbital $\gamma=c$ or $f$, and their Fourier transform, the spin structure factor, 
\begin{equation}
    S^{\gamma}(\textbf{q}) = \frac{1}{L^2} \sum_{\mathbf{i},\mathbf{j}} e^{i (\mathbf{i} - \mathbf{j}) \cdot \textbf{q}} \langle {\bf S}^{\gamma}_{\mathbf{i}} \cdot {\bf S}^{\gamma}_{\mathbf{j}} \rangle,
    \label{Eq2}
\end{equation}
with $L$ being the linear size of the system. The peak of $S^{\gamma}(\textbf{q})$ defines the leading wave-vector $\bf q$ in which the system may exhibit magnetic order, wherein $S^{\gamma}(\textbf{q})\propto L^2$. Given the magnetic structure factor, we can determine the quantum critical points by the magnetic correlation ratio, defined as
\begin{equation}
    R^{\gamma}=1-\frac{S^{\gamma}(\textbf{q}-\delta\textbf{q})}{S^{\gamma}(\textbf{q})},
\end{equation}
where $\textbf{q}=(\pi,\pi)$ and $|\delta\textbf{q}|=\frac{2\pi}{L}$, reflecting the fact that the magnetic order under consideration is of N\'eel type. This quantity reflects the inverse width of the structure-factor peak, which in turn is proportional to the correlation length. Consequently, in the thermodynamic limit, $R^{\gamma}\rightarrow 1$ indicates the emergence of a magnetic order, while $R^{\gamma}\rightarrow 0$ reflects its absence. Since $R^{\gamma}$ is a renormalization group (RG) invariant, the crossing of $R^{\gamma}$ curves for different lattice sizes provides a reliable estimate of the critical point location \cite{Kaul2015, Sato2018, Liu2018}. Moreover, being dimensionless, it obeys the finite-size scaling (FSS) form
\begin{equation}
 R^{\gamma}=\mathcal{F}^{\gamma}[(V-V_c)\,L^{1/\nu}],
  \label{eq:RFSS}
\end{equation}
where $\mathcal{F}^{\gamma}$ is an unknown scaling function and $\nu$ is the correlation length critical exponent. Therefore, $\mathcal{F}^{\gamma}(0)$ is a constant and independent of $L$ at the criticality, such that the curves of $R^{\gamma}$ for different $L$'s cross, and extrapolating these crossing points for $L\to\infty$ provides an estimation for the location of the critical point at the thermodynamic limit. 

\section{Results}
\label{sec:Results}

\subsection{The non-interacting limit}
\label{subsec:Noninteraclim}

Before discussing the emergence of magnetic ordering under strong interactions, we first analyze a non-interacting limit ($U\rightarrow 0$), in which Eq.\,\eqref{eq:Hamiltonian} can be written as a Bloch Hamiltonian,
\begin{equation}
    \mathcal{H}_0=\sum_{\mathbf{k}}\Psi_{\mathbf{k}}^{\dagger}h(\mathbf{k})\Psi_{\mathbf{k}}^{\phantom{\dagger}},
\end{equation}
where the four-component spinor is
\begin{equation}
  \Psi_{\mathbf{k}}^\dagger \;=\;\bigl(c_{\mathbf{k}\uparrow}^\dagger,\;f_{\mathbf{k}\downarrow}^\dagger,\;c_{\mathbf{k}\downarrow}^\dagger,\;f_{\mathbf{k}\uparrow}^\dagger\bigr),
\end{equation}
and
\begin{equation}
  h(\mathbf{k}) \;=\;
  \begin{pmatrix}
    \epsilon_{c}(\mathbf{k})&  V(\mathbf{k}) & 0 & 0 \\[4pt]
    V^*(\mathbf{k}) & \epsilon_{f}(\mathbf{k}) & 0 & 0 \\[4pt]
    0 & 0 & \epsilon_{c}(\mathbf{k})&  V^*(\mathbf{k})\\[4pt]
    0 & 0 & V(\mathbf{k}) & \epsilon_{f}(\mathbf{k}) \\
  \end{pmatrix}.
  \label{eq:Bloch_H}
\end{equation}
Here, the uncoupled dispersion relations of the $c$- and $f$-electrons are described by
\begin{equation}
  \epsilon_{c}(\mathbf{k})
    = -2t\,\bigl(\cos k_x + \cos k_y\bigr)\ ,
    \end{equation}
\begin{equation}
  \epsilon_{f}(\mathbf{k})= \mu_{f}\ ,
\end{equation} 
where $\mu_f$ is the $f$-electrons chemical potential. Additionally, the hybridization term is given by
\begin{equation}
V(\mathbf{k})= -2V\,\bigl(\sin k_x\ + i\sin k_y\bigr).
\end{equation}
From now on, we set $\mu_{f}=0$, and measure all energies in units of $t$. As evident from Eq.~\eqref{eq:Bloch_H}, the $4\times 4$ Bloch Hamiltonian factorizes into two $2\times2$ subblocks, each with eigenvalues,
\begin{equation}
  E_{\pm}(\mathbf{k})
    = \frac{\epsilon_c(\mathbf{k}) + \epsilon_f(\mathbf{k})}{2}
      \pm \sqrt{\left[\dfrac{\epsilon_c(\mathbf{k}) - \epsilon_f(\mathbf{k})}{2}\right]^2 + \epsilon_{V}(\mathbf{k})^2},
\end{equation}
where the hybridization scale is,
\begin{equation}
  \epsilon_{V}(\mathbf{k})
    = 2V\,\sqrt{\sin^2 k_x + \sin^2 k_y}\,.
\end{equation}
Each band $E_{\pm}(\mathbf{k})$ is thus doubly degenerate, reflecting the symmetry associated with the block structure of $h({\bf k})$\,\cite{Legner2014}. 

We now present the resulting band structure and corresponding density of states (DOS). Figure \ref{fig1}(b) shows the case of vanishing hybridization ($V \rightarrow 0$) and nearest-neighbor hopping $t = 1$. In this limit, the $c$- and $f$-orbitals are decoupled: The $f$-band remains completely flat at the Fermi level ($E_F = 0$), while the $c$-band exhibits a dispersive behavior characteristic of a square lattice, with bandwidth $W = 4t$. Consequently, the $f$-orbitals are fully localized, while the $c$-electrons remain itinerant, yielding metallic transport. In this regime, the flat $f$-band—along with the van Hove singularity (vHs) and Fermi surface nesting inherent to the $c$-band—greatly enhances the system's susceptibility to correlation effects once a finite on-site interaction $U$ is introduced.

In the intermediate case with finite hybridization, $V = 0.5$ and $t = 1$ [Fig.~\ref{fig1}(c)], the originally flat $f$-band becomes dispersive due to SOC introduced through the $c$–$f$ hybridization. A hybridization gap opens between the M and $\Gamma$ points in the band structure, while Dirac cones emerge at the X point [$\mathbf{k} = (\pi, 0)$ and its symmetry partner $(0, \pi)$]. Examination of the DOS reveals that the $c$-band behaves as a semimetal, with vanishing spectral weight only at the Fermi level. In contrast, the $f$-band now contributes finite spectral weight at $E_F$, not only from regions such as $\Gamma$ and M with minimal group velocity, but also from the Dirac cone region. In this regime, the vHs is shifted away from the Fermi energy, and the Fermi surface is no longer nested. As a result, correlation effects are suppressed, and a finite critical interaction strength is required to induce magnetic ordering. 

At this point, it is worth noting that the orbital character of each band is illustrated in Fig.\,\ref{fig1} via a colormap representing the difference in orbital projections:
\begin{align}
    P(\mathbf{k}, n) = \left| \bra{f } \ket{\psi_{\mathbf{k}}^n} \right|^2 - \left| \bra{c } \ket{\psi_{\mathbf{k}}^n} \right|^2,
\end{align}
where $\ket{\psi_{\mathbf{k}}^n}$ denotes the $n$-th eigenstate at momentum $\mathbf{k}$, and $\ket{f}$, $\ket{c}$ represent the localized and conduction orbital states, respectively. Using this projection, one observes that while the bands remain mostly unmixed near the $\Gamma$ and $\mathrm{M}$ points, the orbitals become highly hybridized in the vicinity of the Dirac points.

Lastly, in the strong hybridization limit (or `pure SOC'), illustrated in Fig.~\ref{fig1}(d), we set $V = 1$ and take $t \rightarrow 0$. Under these conditions, the bands simplify as $E_{\pm}(\mathbf{k}) = \pm 2V\sqrt{\sin^2 k_x + \sin^2 k_y}$, and the originally localized $f$-band becomes fully dispersive, losing all flat-band characteristics. Simultaneously, the $c$-band acquires the same dispersion, resulting in an identical energy–momentum relation for both orbitals, leading to full hybridization. The band structure exhibits well-defined Dirac cones at the high-symmetry points $\Gamma$, $\mathrm{M}$, and $\mathrm{X}$ of the Brillouin zone, showing that strong SOC hybridization not only delocalizes the $f$-electrons but also equalizes the bandwidths of both orbital species, effectively creating a symmetric, fully itinerant semimetal with two pronounced vHs symmetrically located at $E = \pm 2V$ in the DOS.

\begin{figure}[t]
    \centering
    \includegraphics[width=1\linewidth]{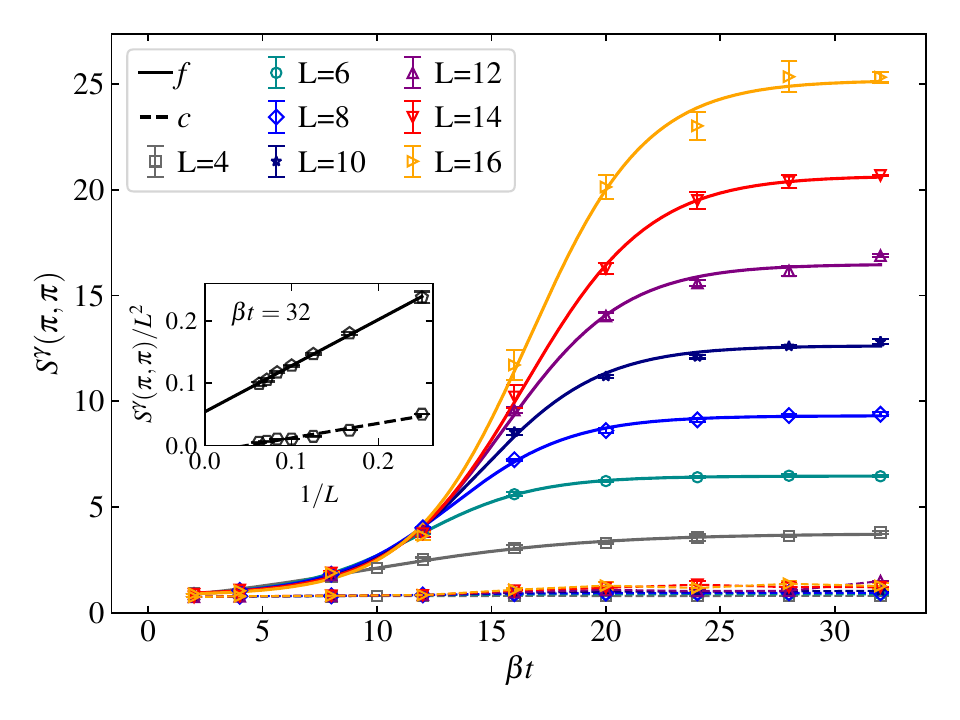} 
    \caption{Magnetic structure factor $S^{\gamma}(\pi,\pi)$ as a function of $\beta t$ for different lattice sizes $L$, under a fixed hybridization $V/t = 0.5$ and on-site electron electron interaction $U/t = 4$. Results for $\gamma = c$- and $f$-orbitals are indicated with dashed and solid lines, respectively. In addition, $\beta$ is the inverse of temperature $T$ ($\beta=1/T$). Inset: Finite-size scaling (FSS) of the magnetic structure factor {$S^{\gamma}(\pi,\pi)$} as a function of $1/L$; a finite thermodynamic extrapolation indicates the existence of a finite order parameter.}
    \label{fig:Spipi_vs_beta}
\end{figure}

In contrast, in the three-dimensional (3D) PAM\,\cite{Legner2014}, the hybridization between $f$- and $c$-electrons splits the spectrum into two bands, without touching at any point. Here, in the two-dimensional (2D) PAM with SOC embedded in the hybridization term, the lattice geometry and SOC combine to yield massless Dirac cones in the band structure. Therefore, in the non-interacting limit, the model realizes a Kondo-Dirac semimetal (KDSM), characterized by the coexistence of Dirac-like quasiparticles and hybridized conduction-localized electron states. The SOC also acts as a key mechanism that suppresses Fermi surface nesting, eliminates flat bands, and shifts the vHs away from half-filling. In the next subsection, we investigate how electronic interactions drive the transition from this KDSM phase into magnetically ordered states.

\subsection{ Strong on-site  electron-electron interaction}

\begin{figure*}[t]
    \centering
    \includegraphics[width=1\linewidth]{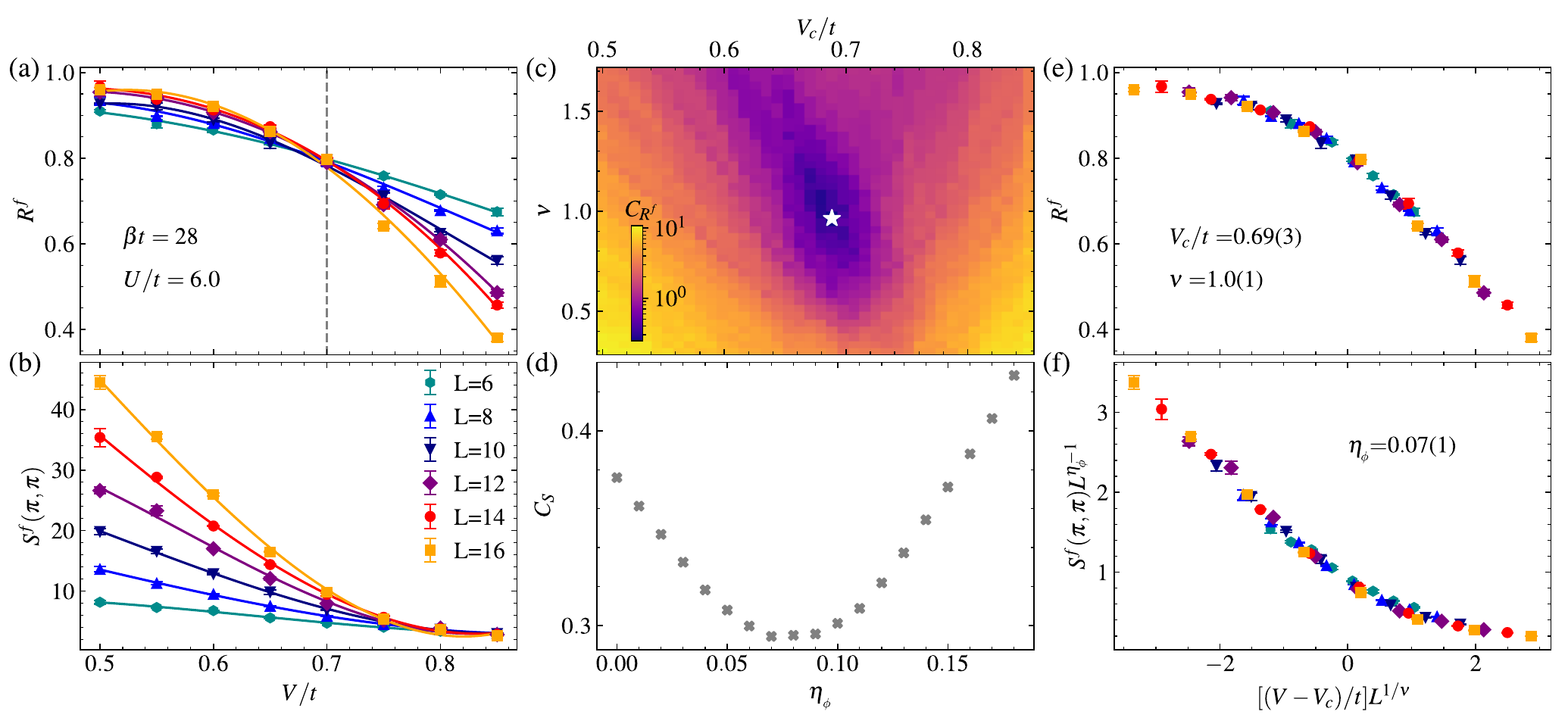}
    \caption{(a) [(b)] The correlation ratio $R^{f}$ [$S^{f}(\pi,\pi)$] as a function of the hybridization term $V$, for different lattice sizes $L$. (c) The colormap of the cost function, $C_{R^f}$, in the $\nu-V_c$ plane. The white star locates its minimum, which in turn determines the best fit to the correlation length exponent $\nu$ and $V_{c}$. (d) The cost function for the structure factor $C_{S^f}$ as a function of the anomalous dimension $\eta_\phi$ at the critical point $V_c$. (e) The resulting collapse of $R^{f}$ with fitted $\nu$ and $V_{c}$ from (c), and (f), the resulting collapse of $S^{f}(\pi,\pi)$  with the anomalous dimension $\eta_\phi$. In all panels, we consider the electron-electron interaction $U/t=6$, and a fixed $\beta t=28$, where $\beta$ is the inverse of temperature $T$ ($\beta=1/T$).
    }
    \label{fig:Collapse}
\end{figure*}

As discussed in the previous section \ref{subsec:Noninteraclim}, any finite SOC in the hybridization term leads to the destruction of nesting, flat bands, and the vHs at \textit{half-filling}. These features tend to suppress the appearance of long-range order in systems with weak interactions. In this section, we will discuss how increasing $U$ can cause a transition to a magnetic phase with indirect coupling between $f$-electrons, mediated by the conduction $c$-orbitals. In our setup, the PAM on the square lattice becomes antiferromagnetically ordered as $U$ increases, mimicking the usual RKKY interaction, even in the presence of spin-flip hybridization terms. 

To quantify that, we show in Fig.\,\ref{fig:Spipi_vs_beta} the antiferromagnetic (AFM) structure factor $S^{\gamma}(\pi,\pi)$ as a function of $\beta t$ (inverse of temperature $T/t$) for different (linear) lattice sizes $L$, and fixed $V/t = 0.5$ and $U/t=4$ (as before, $\gamma=c, f$ labels the two orbital species). Here, $S^{c}(\pi,\pi)$ exhibits little dependence on either $\beta$ or $L$, contrary to what happens to the $f$-electrons, in which $S^{f}(\pi,\pi)$ becomes larger as we increase both lattice size $L$ and $\beta$. At sufficiently low temperatures (here, for $\beta t=32$), $S^{f}(\pi,\pi)$ saturates to a maximum value, denoting that the correlation length is comparable with $L$. To probe the existence of a long-range magnetic order in the thermodynamic limit, we extrapolate $S^{\gamma}(T\to 0)/L^2$ as a function of $1/L$, as displayed in the inset of Fig.\,\ref{fig:Spipi_vs_beta}. It is then clear that there is an absence of long-range magnetic order given by $c$-electrons, while the $f$-orbital presents a finite order parameter in the thermodynamic limit. Therefore, we have evidence that the $f$-electrons, and only they, support the presence of long-range magnetic order.

In addition, while $V$ regulates the magnitude of the SOC, it acts as an effective RKKY mechanism capable of inducing magnetic order in the localized electrons. Yet, it does not lead to nontrivial magnetic textures, such as those reported in the Rashba-Hubbard model\,\cite{SousaJunior2025}, where, in particular, the $S^{f}(\pi,\pi)$ consistently exhibits its maximum at the wave vector $\qv = (\pi,\pi)$, indicating the stabilization of a conventional antiferromagnetic order. That is, the SOC is unable to induce any anisotropy in the spin correlations, thereby preserving SU$(2)$ spin-rotation symmetry. Although this symmetry is explicitly broken at the level of individual configurations due to the simple form of the HS transformation used here\,\cite{Hirsch1983}, it is statistically restored through the stochastic sampling procedure employed in the DQMC simulations.

As we have shown in the non-interacting limit (see Sec.\,\ref{subsec:Noninteraclim} for further details), increasing $V/t$ enhances the Fermi velocity and renders the $f$-band more dispersive, thereby suppressing magnetic ordering. Consequently, for any finite value $V/t$, there exists a critical value of $U_{c}$ at which the model undergoes a phase transition between the KDSM and AFM phases. This quantum phase transition is examined in greater detail in Fig.\,\ref{fig:Collapse}, where we fix $U/t = 6$ and $\beta t = 28$, corresponding to a sufficiently low temperature to capture ground-state properties for this parameter set reliably. As expected, for small $V/t$, both $R^{f}$ and $S^{f}(\pi,\pi)$ increase with $L$ as illustrated in Figs.\,\ref{fig:Collapse}(a) and \ref{fig:Collapse}(b), respectively. Still, for $V/t \gtrsim 0.7$ (vertical dashed line). $R^f$ starts to decrease with $L$, and the $S^{f}(\pi,\pi)$ does not acquire a significant system size dependence beyond this point. 

To more precisely estimate the quantum critical point, we perform an FSS analysis. For this purpose, we employ the cost function $C_y = \sum_j (|y_{j+1} - y_j|)/(\max\{y_j\} - \min\{y_j\})-1$\,\cite{Suntajs2020, Jin2022, Mondaini2023}, where $y_j$ represents the values of $R^{f}$ [or $S^{f}(\pi,\pi)$] ordered according to their rescaled variable $[(V-V_c)/t]L^{1/\nu}$. $C_{y}$ thus quantifies the deviation between consecutive points in the scaling collapse: its minimum determines the set of parameters that optimizes a smooth dependence on the rescaled variable. The cost function values computed for $R^{f}$, and denoted by $C_{R^f}$, are shown in Fig.\,\ref{fig:Collapse}(c). The optimal values of $\nu$ and $V_c$ that minimize $C_{R^f}$ are marked with a star, and using these parameters one obtains the collapsed $R^{f}$ in Fig.\,\ref{fig:Collapse}(e). The FSS analysis of $R^{f}$ yields the correlation-length exponent $\nu = 1.0(1) $ and critical hybridization $V_c/t=0.69(3)$, for this interaction strength. 

Fixing these values, we further estimate the bosonic anomalous dimension $\eta_\phi$, governing the finite-size scaling of the antiferromagnetic order parameter and its structure factor. At criticality, it scales as $S^f(\pi,\pi)\sim L^{d-z-\eta_\phi}$~\cite{Sandvik2010}, where $d=2$ is the dimension, and $z=1$ is the dynamic critical exponent for a relativistic-like scaling, so that $S^f(\pi,\pi) L^{\eta_\phi-1} = g([(V-V_c)/t]L^{1/\nu})$, as illustrated in Fig.\,\ref{fig:Collapse}(f). This is done by minimizing the associated $C_{y}$ computed from $S^{f}(\pi,\pi)$, which we denote as $C_S$, and display in Fig.\,\ref{fig:Collapse}(d), and whose minimum provides an estimate of $\eta_\phi$.

These results suggest a scenario in which Dirac electrons, driven by SOC, form hybridized orbitals, resulting in a linear energy dispersion near the Fermi level. In this context, the quantum phase transition into the AFM phase (here at sufficiently strong $V$) is commonly associated with the Gross-Neveu universality class\,\cite{Assaad2013, Parisen2015, Otsuka2016, Tang2018, Otsuka2020}. The corresponding correlation-length exponent reported for this class is in good agreement with those obtained in our analysis, namely $\nu = 1.0(1)$. Remarkably, in Dirac structures, even when non-local interactions are included, the correlation length critical exponent remains close to $\nu \sim 1$\,\cite{Tang2018, Kennedy2025}. The anomalous dimension has also been evaluated in Refs.\,\cite{Otsuka2016, Otsuka2020} for the honeycomb and \(\pi\)-flux Hubbard models, which realize the same Dirac semimetal-to-AFM universality class. A quantitative comparison of $\eta_\phi$ across different lattice realizations is, however, more delicate than for the correlation-length exponent $\nu$: while $\eta_\phi$ is universal in principle, its numerical estimates exhibit a larger spread and stronger corrections to scaling, in particular because subleading corrections $\propto L^{-\omega}$~\cite{Sandvik2010, Otsuka2020} are often required for a fully quantitative extraction. In contrast, $\nu$ appears more consistent among different models.

\begin{figure}[t]
    \centering
    \includegraphics[scale= 0.5]{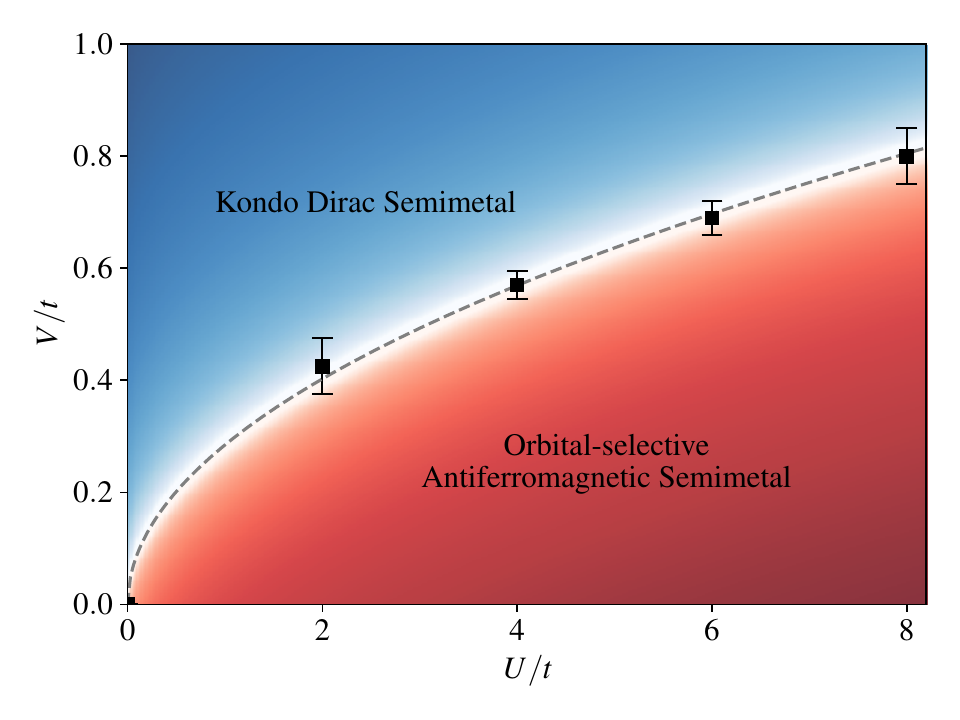}
    \caption{The ground state phase diagram of spin-orbit-coupled (SOC) hybridization $V$ versus on-site electron-electron correlation $U$ in the two-dimensional (2D) periodic Anderson model (PAM). Data points are obtained by compiling the crossing points of the correlation ratio $R^{f}$ vs.~$V/t$, as in Fig.\,\ref{fig:Collapse}(a) for different interaction strengths. The dashed line serves as a guide to the eye, inspired by the mean-field results (see Appendix \ref{App:MFT}).}
    \label{fig:phase_diagram}
\end{figure}

\begin{figure*}[t]
    \centering
    \includegraphics[scale = 0.5]{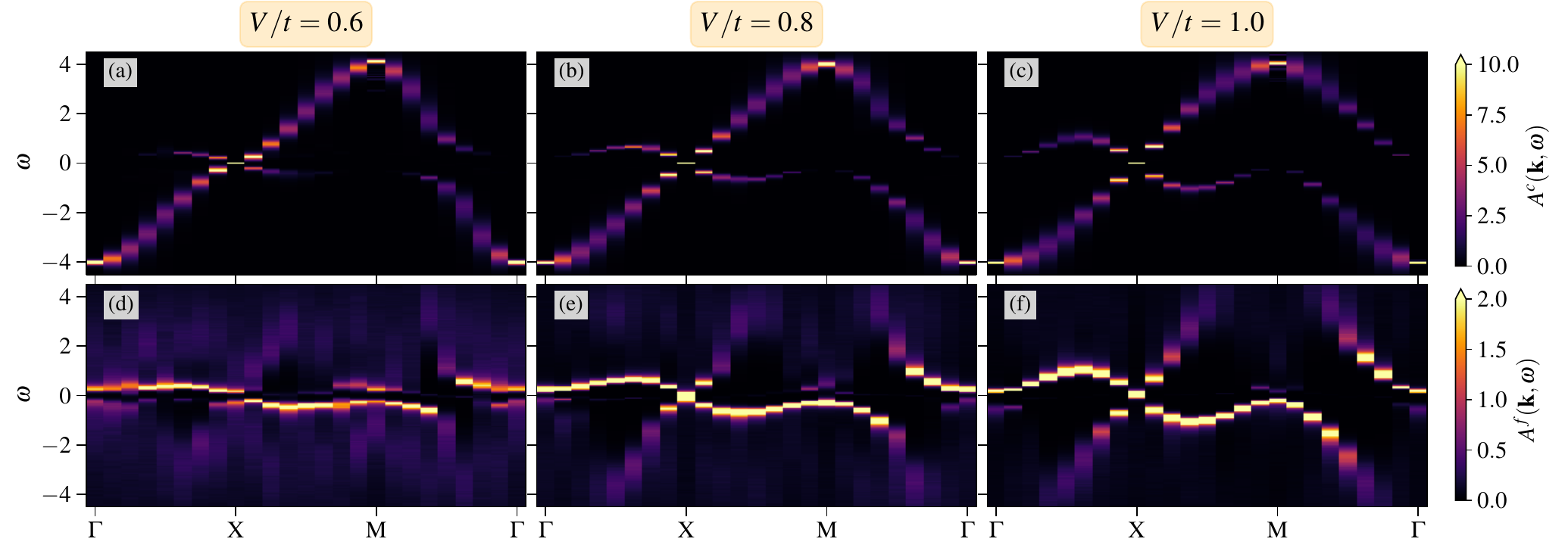}
    \caption{The spectral functions $A^\gamma({\bf k},\omega)$ for both $c$- and $f$-orbitals. In (a), the hybridization with spin-orbit coupling is $V/t=0.6$, in (b) $V/t=0.8$, and in (c) $V/t=1.0$. We consider a fixed on-site electron-electron interaction $U/t=6$ for the $f$-orbitals [the same as in Fig.~\ref{fig:Collapse}]; linear lattice size is $L=16$, and $\beta t = 28$.
    }
    \label{fig:Akw_U6}
\end{figure*}

Extending these results, particularly those based on the correlation ratio, it is possible to estimate the critical point $V_c(U)$ at which the quantum phase transitions occur by collecting the crossing points of $R^{f}$, as in Fig.\,\ref{fig:Collapse}(a), for different interaction strengths. Based on this analysis, one can compile the ground-state phase diagram in the $V/t$ versus $U/t$ plane, as shown in Fig.\,\ref{fig:phase_diagram}. The diagram summarizes the parameter regimes in which the interplay between magnetic ordering and Dirac electrons gives rise to two distinct phases. For increasing $V/t$, hybridization becomes the dominant energy scale, and the presence of SOC favors the formation of Dirac cones. In this regime, the local magnetic moments in the $f$-orbitals do not exhibit long-range order, leading to the emergence of a KDSM phase, characterized by strongly hybridized quasiparticles and screened magnetic correlations.

In contrast, in the weak $V/t$ regime, the interaction $U/t$ enhances the AFM correlations, ultimately stabilizing a long-range AFM phase. This behavior is consistent with the RKKY mechanism, wherein the local moments interact via the itinerant $c$-electrons~\cite{Coleman2007}. In both phases, our results suggest that the system retains semimetallic behavior, at least of an orbitally selective type, as will become clear in the following discussion. Indeed, for further characterization of the electronic structure of these phases, we compute the spectral weight functions in the next subsection.

\subsection{Spectral properties}
\label{subsec:Spectral_Akw}

To analyze how the electron-electron interactions renormalize the electronic spectrum, we compute the spectral function $A^{\gamma}(\textbf{k},\omega)$ for both $\gamma=c,f$-orbitals. For this purpose, for each momentum $\kv$ and orbital $\gamma$, we first extract the imaginary-time Green's functions $G^{\gamma}_{\bf k}(\tau)$, and use the standard finite-temperature fermionic kernel $K(\tau,\omega)=\frac{e^{-\omega \tau}}{1+e^{-\beta\omega}}$ to perform the analytic continuation to real frequencies:
\begin{equation}
G^{\gamma}_{\bf k}(\tau) =  \int d{\omega} K(\tau,\omega) A^{\gamma}({\bf k},\omega)\ .
\end{equation} 
This is achieved by using the stochastic Maximum Entropy implementation provided in the ALF package~\cite{Assaad2022}, which enables the reconstruction of real-frequency spectral functions and allows us to investigate both interaction-induced gap opening and the renormalization of the Fermi velocity. 

Figure \ref{fig:Akw_U6} displays the spectral function $A^\gamma(\textbf{k},\omega)$ for both $c$- [Figs.\,\ref{fig:Akw_U6}(a), (c), and (e)] and $f$- [Figs. \ref{fig:Akw_U6}(b), (d), and (f)] orbitals under $U/t=6$ and $\beta t = 28$. For this set of parameters, $V_c/t \simeq 0.69$ [see Fig.~\ref{fig:Collapse}], and at a value smaller than that, e.g., $V/t=0.6$, the system exhibits AFM with a finite staggered magnetization. Indeed, at this hybridization level, Figs.\,\ref{fig:Akw_U6}(a) and \ref{fig:Akw_U6}(d) show an incipient, but finite, Mott gap for the $f$-orbital, while the $c$-orbital remains in a semimetallic state, since the spectral weight at the Fermi level arises only from the Dirac point. A comprehensible analysis of gap formation in the presence or absence of SOC in the $c$-$f$ hybridization is provided in Appendix \ref{App:pam_wsoc}. 

Increasing $V/t$ to 0.8, mapping a regime just outside the AFM phase, Figs.\,\ref{fig:Akw_U6}(b) and \ref{fig:Akw_U6}(e) show that the gap in the $f$-orbital vanishes, signaling the absence of order, while the $c$-orbitals still exhibit the characteristic semimetallic $A^{c}(\textbf{k},\omega)$ profile. At even larger hybridization, $V/t=1.0$, Figs.\,\ref{fig:Akw_U6}(c) and \ref{fig:Akw_U6}(f) show the spectral weight corresponding to the KDSM phase, where the absence of a gap for excitations in both  $c$- and $f$-orbitals arises. In fact, $A^{c}(\textbf{k},\omega)$ and $A^{f}(\textbf{k},\omega)$  closely resemble the band structure of the non-interacting limit, previously shown in Fig.\,\ref{fig1}(c) in Sec.\,\ref{subsec:Noninteraclim}. The difference stems from the $\Gamma$ and M points, where the spectral weight is pushed away from the Fermi level, such that only the Dirac point contributes to a finite spectral weight in the limit $\omega \to 0$, characterizing a correlated semimetallic state.

Since the interaction $U/t$ acts exclusively on the $f$-orbitals, the Mott transition is orbital-selective, a feature commonly observed in Kondo lattice systems with non-local hybridization\,\cite{Zhang2019}, and here evidenced by the $c$-orbitals remaining gapless across all interaction regimes analyzed. Moreover, the Dirac cone structure at the X point is preserved, with only a slight renormalization of the Fermi velocity. This renormalization stems from the localized AFM order in the $f$-electrons, which partially suppresses the itinerancy of the $c$-electrons, especially near the Dirac points. In contrast, the $c$-orbital spectral function $A^{c}(\textbf{k},\omega)$ remains largely unmodified at the $\Gamma$ and M points, where the spectral peaks are numerically indistinguishable from those in the non-interacting limit.


\section{Summary and discussion}
\label{sec:Conclusions}

We investigate the Mott transition and the emergence of antiferromagnetic long-range order in a system where a conduction orbital hybridizes with a localized one through a spin-orbit coupled term. By tuning the interaction strength, we estimate the critical value of hybridization that destroys the magnetic order. With this, we establish the entire ground state phase diagram by extrapolating the results to both the zero-temperature and thermodynamic limits. As the spin-orbit coupling gives rise to Dirac cones at the $\rm{X}$ point of the Brillouin zone, we also check the compatibility of the critical exponents of the phase transition with the Gross-Neveu-Heisenberg universality class, which is typical of Dirac fermion systems that exhibit gap opening stemming from a spontaneous SU(2) symmetry breaking. Our numerical results have shown that the critical exponents indeed fit those found for this universality class. As the Hubbard interaction directly couples only the $f$-electrons, we found the Mott transition to be orbital-selective, where only the spectrum of the $f$-electrons is gaped, while the Dirac cones in the conduction survive for any value of interaction strength. 

The robustness of the Dirac cones within the magnetic phase is particularly noteworthy due to their potential topological implications. A recent study of the antiferromagnetic Dirac semimetal candidate \ce{GdIn_3} highlighted how magnetic order can coexist with Dirac fermions~\cite{Yin2022}. In the three-dimensional periodic Anderson model, non-trivial topological phases have been reported~\cite{Dzero2010,Dzero2012,Dzero2016,Legner2014}, and similar behavior has also been proposed in a modified two-dimensional setting~\cite{Luo2021}. The competition between topological and magnetic orders remains an open question~\cite{Masuda2016,Li2019, Li2020, Yang2020, Klett2020, Ido2024}, and our results provide an additional example of how Dirac-like quasiparticles, Kondo hybridization, and antiferromagnetic order can coexist and compete, thereby contributing to a deeper understanding of this interplay. 

While Monte Carlo simulations have addressed the standard three-dimensional periodic Anderson model in the absence of spin–orbit coupling~\cite{Oliveira2023}, extending these studies to include spin–orbit interaction in three dimensions is a compelling direction for future research. Notably, the regime in which the $f$-electrons possess a finite hopping $t_f$, such that $t_{f}/t<0$, is expected to host non-trivial topological phases and is also free from the sign problem in determinant quantum Monte Carlo simulations. Based on our findings, we conclude that the periodic Anderson model with spin–orbit coupling provides a versatile and rich platform for exploring interaction-driven phenomena in correlated Dirac and heavy-fermion systems. Lastly, the gapless nature of the conduction $c$-band can significantly affect the transport properties of Kondo systems, especially in doped regimes, where it may also favor unconventional pairing mechanisms.

\section*{Acknowledgments}
S.A.S.-J. gratefully acknowledges the Brazilian agency CNPq for funding support, grant No.~201000/2024-5. J.~F. acknowledges support from ANID Fondecyt grant number 3240320. R.M.~acknowledges support from the T$_{\rm c}$SUH Welch Professorship Award. Powered@NLHPC: This research was partially supported by the supercomputing infrastructure of the NLHPC (CCSS210001). Numerical simulations were performed with resources provided by the Research Computing Data Core at the University of Houston. This work also used TAMU ACES at Texas A\&M HPRC through allocation PHY240046 from the Advanced Cyberinfrastructure Coordination Ecosystem: Services \& Support (ACCESS) program, which is supported by U.S. National Science Foundation grants 2138259, 2138286, 2138307, 2137603, and 2138296. The data that support the findings of this article are openly available~\cite{zenodo}.


\appendix

\section{Sign Problem}
\label{App:sign_prob}

In this Appendix, we demonstrate the absence of the sign problem by employing a combination of particle-hole and time reversal transformations. 
Firstly, we note that the hopping part of the Hamiltonian can be blocked using the following spinor basis, $\Psi^\dagger = (c_\uparrow^\dagger \, f_\downarrow^\dagger \, c_\downarrow^\dagger \, f_\uparrow^\dagger )$, similar to the process done in Sec.\,\ref{subsec:Noninteraclim}. Suppressing site indices for simplicity, the real-space hopping Hamiltonian of the system can be cast as,
\begin{align}
\mathcal{H}_0 = \Psi^\dagger
\begin{pmatrix}
A & 0 \\
0 & B
\end{pmatrix}
\Psi\ ,
\end{align}
where $A$ and $B$ are  are single-particle hopping matrices acting in the $(c_\uparrow,f_\downarrow)$ and $(c_\downarrow,f_\uparrow)$ sectors, respectively. A sufficient condition for a sign-problem-free DQMC simulation is that, for every configuration of the auxiliary fields, these blocks satisfy $B = A^*$, so that the fermionic weight factorizes as $\det A \,\det B = |\det A|^2 \ge 0$.

\begin{figure}[t]
    \centering
    \includegraphics[scale = 0.5]{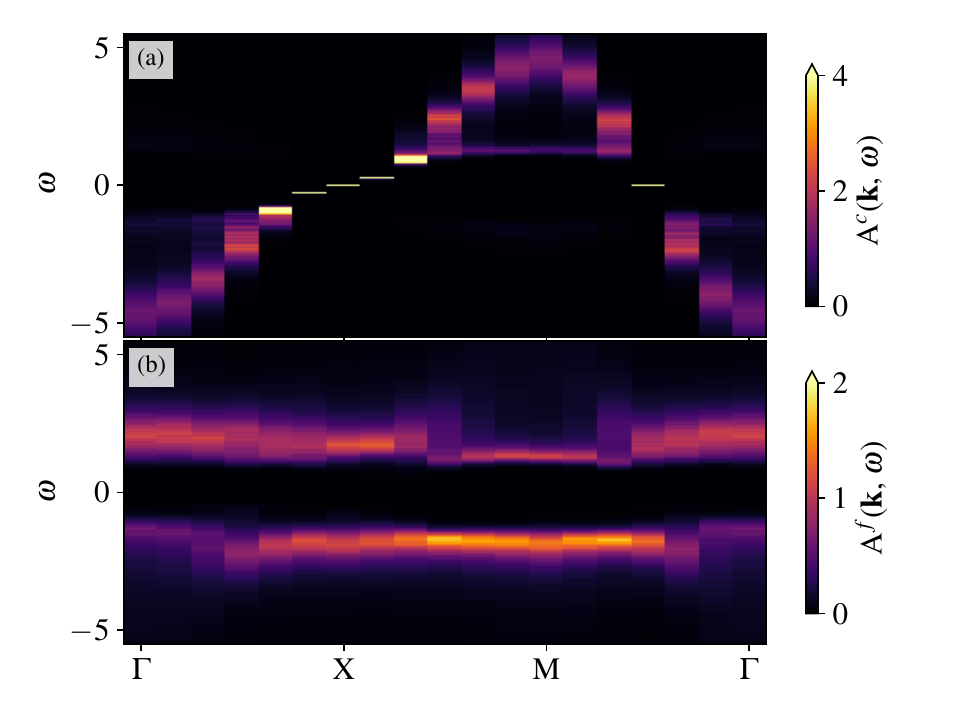}
    \caption{Spectral function $A^\gamma(\textbf{k},\omega)$ of the periodic Anderson model (PAM) for (a) $c$- and (b) $f$-orbitals. Here we set the Coulomb interaction $U/t=5$, inverse temperature $\beta t = 14$, lattice size $L=12$, and the hybridization term as $V_0/t=0.5$, in the absence of spin-orbit coupling (SOC). This model follows the definition given in Ref.\,\cite{Zhang2019}.}
    \label{fig:pam_wo_soc}
\end{figure}

To get this relation, we apply a particle-hole transformation on the $B$ block, as illustrated in Ref.\,\cite{Zheng2011},
\begin{align}
    B_{m,n} \rightarrow & B_{m,n}^*,  
\end{align}
where $B_{m,n}$ are the matrix elements of the block $B$ matrix. In sequence, we also make a time-reversal transformation,
\begin{align}
\mathcal{T} \alpha_{i\uparrow} \mathcal{T}^{-1} &= \alpha_{i\downarrow}, &
\mathcal{T} \alpha_{i\downarrow} \mathcal{T}^{-1} &= -\alpha_{i\uparrow}, \\
\mathcal{T} \alpha_{i\uparrow}^\dagger \mathcal{T}^{-1} &= \alpha_{i\downarrow}^\dagger, &
\mathcal{T} \alpha_{i\downarrow}^\dagger \mathcal{T}^{-1} &= -\alpha_{i\uparrow}^\dagger. \\
\mathcal{T} B_{m,n} \mathcal{T}^{-1} &= B_{m,n}^*.
\end{align}
Here, $\alpha = {c, f}$ denotes the fermionic operators. Under the applied transformations, the matrix blocks satisfy the desired relation: one block is the complex conjugate of the other. It is important to emphasize that the $c \leftrightarrow f$ hopping occurs between different sublattices. This structure enables the determinant of the entire matrix to be expressed as the product of two complex-conjugate determinants, ensuring that the total determinant is always real and positive. 

This rearrangement of the basis, which casts the Hamiltonian into block form, not only makes the absence of the sign problem explicit but also enhances computational efficiency in DQMC simulations, as the matrix dimensions are effectively reduced by half. Furthermore, since the block structure relies on particle-hole and time-reversal gauge transformations, certain constraints are imposed on the Hamiltonian, such as setting $\mu_{f}=0$ and excluding the possibility of any external Zeeman field. If, for example, Rashba spin-orbit coupling or orbital hybridization without spin-flip processes is included [see Appendix \ref{App:pam_wsoc}], this block decomposition is no longer possible, and the sign problem indeed emerges. Finally, we note that turning on the hopping amplitude $t_f$ for the $f$-electrons in Eq. \eqref{eq:Hamiltonian} still preserves the sign-problem-free nature of the model. 

\section{Spectral gaps}
\label{App:pam_wsoc}

\begin{figure}[t]
    \centering
    \includegraphics[scale = 0.5]{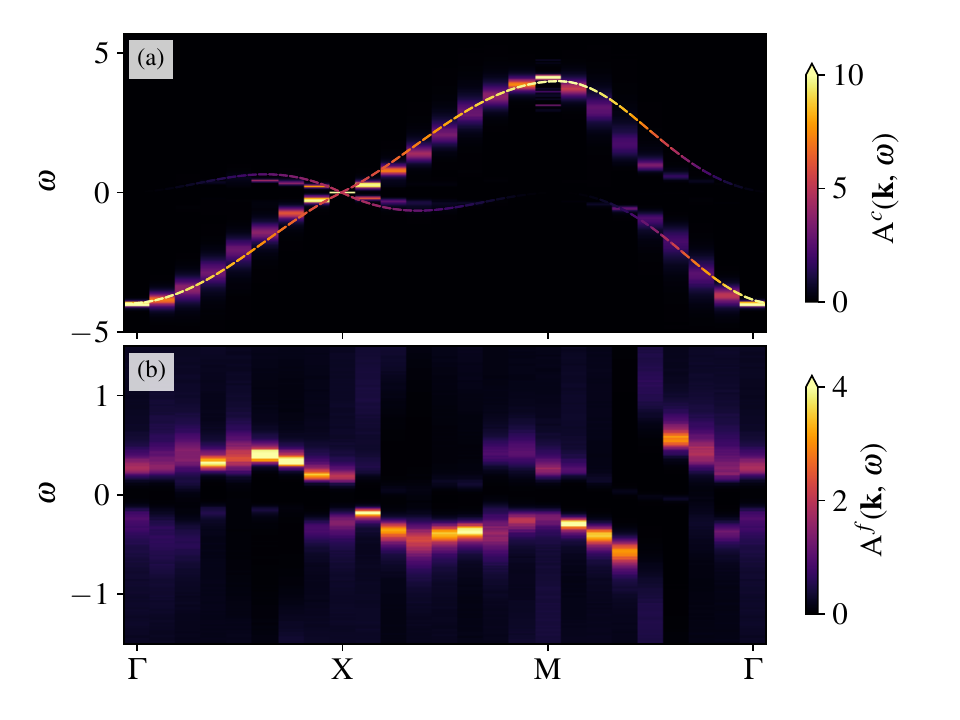}
    \caption{Spectral function $A^\gamma(\textbf{k},\omega)$ of the periodic Anderson model (PAM) for (a) $c$- and (b) $f$-orbitals. Here we set the Coulomb interaction $U/t=6$, inverse temperature $\beta t = 16$, lattice size $L=16$, and the hybridization term $V/t=0.6$, in the presence of spin-orbit coupling (SOC). These results are the same as the ones in Fig.\,\ref{fig:Akw_U6}(a) and \ref{fig:Akw_U6}(d), but the frequency $\omega$-axis has been rescaled to highlight better the gap region. The dashed lines in (a) show the spectral function in the non-interacting ($U/t=0$) case for comparison.}
    \label{fig:pam_with_soc}
\end{figure}

In this Appendix, we expand on the discussion of the orbital-selective gap opening in the system and the emergence of Dirac cones through the inclusion of SOC in the hybridization term of the PAM. To better compare our present case with previous studies of the PAM, we calculate the spectral function for the Hamiltonian analyzed in Ref.\,\cite{Zhang2019}
\begin{align}
\mathcal{H}_{V} = V_0\sum_{\langle \iv,\jv \rangle} \!\!\left[ c_\iv^\dagger f^{}_\jv\!+ f_\iv^\dagger c_\jv\!+\!{\rm H.c.}\right]\ .
\label{H_hib_wo_soc}
\end{align}
which does not feature SOC for the hybridization term, and if taking $V\rightarrow V_{0}$, all definitions of every operator are equivalent to those in Sec.\,\ref{sec:Method}. It is important to note that this model remains sign problem-free only when either $V$ or $V_0$ is finite, but not both simultaneously. 

As reported in Ref.\,\cite{Zhang2019}, when  SOC is absent, a large $V_{0}$ produces a metallic singlet phase, although no Dirac cone features appear. Conversely, in the AFM phase, a pronounced gap develops in the spectrum of the $f$-orbital, whereas the $c$-orbital retains metallic character. To illustrate this behavior, in Fig.\,\ref{fig:pam_wo_soc} we show $A(\textbf{k},\omega)$ fo $c$-orbitals and [Fig.\,\ref{fig:pam_wo_soc}(a)] $f$-orbitals [Fig.\,\ref{fig:pam_wo_soc}(b)] under fixed $L=12$, $U/t=5$, and $V_0/t=0.5$ within the AFM phase. As in the main text, the spectral functions $A^\gamma(\textbf{k},\omega)$ were obtained using the stochastic maximum entropy method\,\cite{Assaad2022}. The results indicate that the behavior of $A^\gamma(\textbf{k},\omega)$ is analogous to that found in Ref.\,\cite{Zhang2019}, although we interpret them from the perspective of band structures.

To compare with the previous results obtained in the absence of SOC, we now analyze the scenario in which the Hamiltonian, Eq.\,\eqref{eq:Hamiltonian} includes SOC in $V$. Fig. ~\ref{fig:pam_with_soc} shows $A(\mathbf{k},\omega)$ for (a) $c$-orbitals and (b) $f$-orbitals, under the fixed $L=16$, $U/t=6$, and $V/t=0.6$, as in Fig.~\ref{fig:Akw_U6}. These values place the system inside the AFM region of the ground-state phase diagram shown in Fig.~\ref{fig:phase_diagram}. 

Our results exhibit behavior similar to that reported in Ref.~\cite{Zhang2019}, where the $c$-orbitals remain (semi-)metallic, while the $f$-orbitals develop a well-defined insulating gap in the AFM phase. This suggests that the opening of the gap in the $f$-orbitals is not a direct consequence of the SOC in $V$, but rather an intrinsic feature of the PAM in the correlated AFM regime with a nonlocal $V$. The primary effect of SOC in our model is to create Dirac cones, thereby inducing a semimetallic state.

\begin{figure}[t]
\centering
\includegraphics[scale = 0.5]{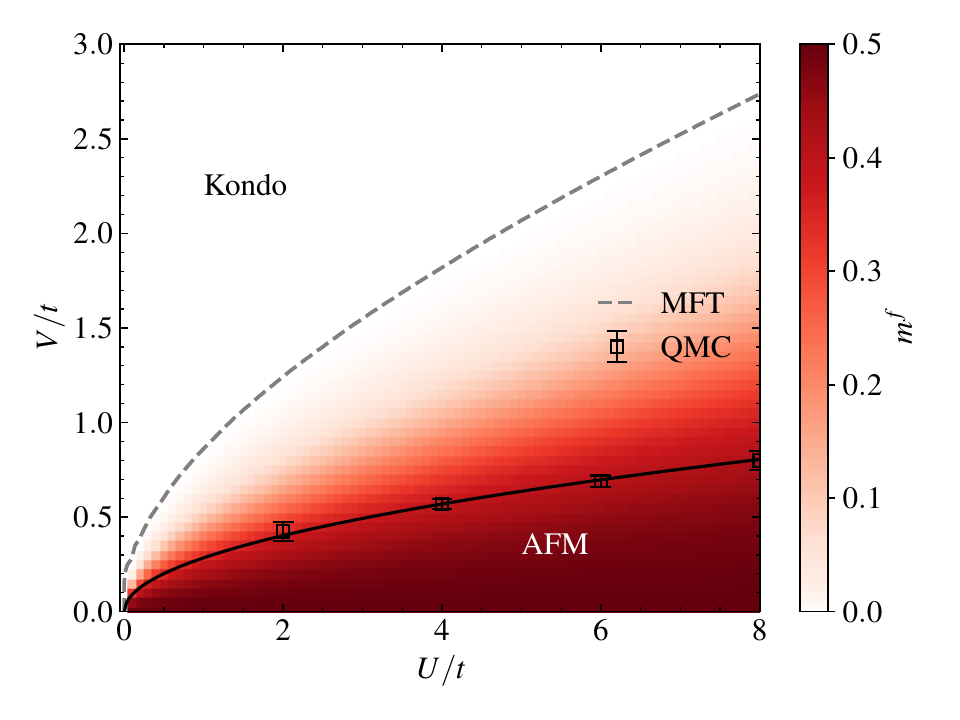}
\caption{The ground state mean-field phase diagram of spin-orbit-coupled hybridization ($V/t$) versus on-site electron-electron correlation ($U/t$) in the Periodic Anderson model. The color scale indicates the magnitude of the mean-field order parameter, while the dashed line marks the threshold $m^f \sim 10^{-3}$ used to define the phase boundary. The markers represent the determinant quantum Monte Carlo critical points [see Fig.\,\ref{fig:phase_diagram}], and the solid line serves as a guide to the eye.}
\label{fig:MFT}
\end{figure}

\section{Mean-Field Approximation}
\label{App:MFT}

To gain an initial understanding of the magnetic phase behavior across a broad region of the parameter space, we first apply a Hartree-Fock approximation to the PAM. Within this framework, and considering only the magnetic ansatz, the Hubbard interaction can be decoupled as
\begin{align}
\sum_\iv \left(n^f_{\iv\uparrow}-\frac{1}{2}\right)\left(n^f_{\iv\downarrow}-\frac{1}{2}\right)
\approx \sum_{\iv} \left[ - 2\,\mathbf{m}^f_{\iv}\cdot \mathbf{S}_{\iv} + (m^f)^2 \right],
\end{align}
where the density-dependent contribution can be neglected at \textit{half-filling}. The local magnetization is defined as
\begin{equation}
\mathbf{m}^f_{\iv} = \langle \mathbf{S}_{\iv} \rangle = m^f[\cos(\mathbf{q}\cdot \iv), \sin(\mathbf{q}\cdot \iv), 0],
\end{equation}
with
\begin{equation}
\mathbf{S}_{\iv} = \frac{1}{2}\sum_{\s,\s'=\uparrow,\downarrow }
c^{\dagger}_{\iv\s}\pmb{\sigma}_{\s\s'}c_{\iv\s'}.
\end{equation}
Guided by the DQMC results, we adopt the Néel ordering vector $\mathbf{q} = (\pi,\pi)$. After transforming the mean-field Hamiltonian into momentum space, the order parameter is determined self-consistently.

The resulting order parameter $m^f$ is shown in Fig.~\ref{fig:MFT}. At the mean-field level, the approach captures both phases and qualitatively reproduces the shape of the transition line. However, as is typical for Hartree-Fock treatments, the extent of the magnetically ordered phase is overestimated~\cite{Hu2017}. This behavior also occurs in the PAM with SOC, where mean-field calculations predict a significantly larger AFM region compared to DQMC results, as seen in Fig.~\ref{fig:MFT}. In addition to overestimating the ordered phase, the mean-field approach fails to yield reliable spectral properties: the AFM phase lacks the Dirac cone features observed in the DQMC spectra, and the renormalization of the Kondo bands is also absent. These discrepancies underscore the importance of the DQMC approach—not only for accurately determining the phase boundary, but also for correctly capturing the spectral properties of the structure system.

\bibliography{references}

\begin{thebibliography}{79}%
\makeatletter
\providecommand \@ifxundefined [1]{%
 \@ifx{#1\undefined}
}%
\providecommand \@ifnum [1]{%
 \ifnum #1\expandafter \@firstoftwo
 \else \expandafter \@secondoftwo
 \fi
}%
\providecommand \@ifx [1]{%
 \ifx #1\expandafter \@firstoftwo
 \else \expandafter \@secondoftwo
 \fi
}%
\providecommand \natexlab [1]{#1}%
\providecommand \enquote  [1]{``#1''}%
\providecommand \bibnamefont  [1]{#1}%
\providecommand \bibfnamefont [1]{#1}%
\providecommand \citenamefont [1]{#1}%
\providecommand \href@noop [0]{\@secondoftwo}%
\providecommand \href [0]{\begingroup \@sanitize@url \@href}%
\providecommand \@href[1]{\@@startlink{#1}\@@href}%
\providecommand \@@href[1]{\endgroup#1\@@endlink}%
\providecommand \@sanitize@url [0]{\catcode `\\12\catcode `\$12\catcode `\&12\catcode `\#12\catcode `\^12\catcode `\_12\catcode `\%12\relax}%
\providecommand \@@startlink[1]{}%
\providecommand \@@endlink[0]{}%
\providecommand \url  [0]{\begingroup\@sanitize@url \@url }%
\providecommand \@url [1]{\endgroup\@href {#1}{\urlprefix }}%
\providecommand \urlprefix  [0]{URL }%
\providecommand \Eprint [0]{\href }%
\providecommand \doibase [0]{https://doi.org/}%
\providecommand \selectlanguage [0]{\@gobble}%
\providecommand \bibinfo  [0]{\@secondoftwo}%
\providecommand \bibfield  [0]{\@secondoftwo}%
\providecommand \translation [1]{[#1]}%
\providecommand \BibitemOpen [0]{}%
\providecommand \bibitemStop [0]{}%
\providecommand \bibitemNoStop [0]{.\EOS\space}%
\providecommand \EOS [0]{\spacefactor3000\relax}%
\providecommand \BibitemShut  [1]{\csname bibitem#1\endcsname}%
\let\auto@bib@innerbib\@empty
\bibitem [{\citenamefont {Coleman}(2007)}]{Coleman2007}%
  \BibitemOpen
  \bibfield  {author} {\bibinfo {author} {\bibfnamefont {P.}~\bibnamefont {Coleman}},\ }\bibfield  {title} {\bibinfo {title} {Heavy fermions: Electrons at the edge of magnetism},\ }\href {https://doi.org/10.1002/9780470022184.hmm105} {\bibfield  {journal} {\bibinfo  {journal} {Handbook of Magnetism and Advanced Magnetic Materials}\ }\textbf {\bibinfo {volume} {1}},\ \bibinfo {pages} {95} (\bibinfo {year} {2007})}\BibitemShut {NoStop}%
\bibitem [{\citenamefont {Kimura}\ \emph {et~al.}(2005)\citenamefont {Kimura}, \citenamefont {Ito}, \citenamefont {Saitoh}, \citenamefont {Umeda}, \citenamefont {Aoki},\ and\ \citenamefont {Terashima}}]{Kimura2005}%
  \BibitemOpen
  \bibfield  {author} {\bibinfo {author} {\bibfnamefont {N.}~\bibnamefont {Kimura}}, \bibinfo {author} {\bibfnamefont {K.}~\bibnamefont {Ito}}, \bibinfo {author} {\bibfnamefont {K.}~\bibnamefont {Saitoh}}, \bibinfo {author} {\bibfnamefont {Y.}~\bibnamefont {Umeda}}, \bibinfo {author} {\bibfnamefont {H.}~\bibnamefont {Aoki}},\ and\ \bibinfo {author} {\bibfnamefont {T.}~\bibnamefont {Terashima}},\ }\bibfield  {title} {\bibinfo {title} {Pressure-induced superconductivity in noncentrosymmetric heavy-fermion {CeRhSi}$_{3}$},\ }\href {https://doi.org/10.1103/PhysRevLett.95.247004} {\bibfield  {journal} {\bibinfo  {journal} {Phys. Rev. Lett.}\ }\textbf {\bibinfo {volume} {95}},\ \bibinfo {pages} {247004} (\bibinfo {year} {2005})}\BibitemShut {NoStop}%
\bibitem [{\citenamefont {Pfleiderer}(2009)}]{Pfleiderer2009}%
  \BibitemOpen
  \bibfield  {author} {\bibinfo {author} {\bibfnamefont {C.}~\bibnamefont {Pfleiderer}},\ }\bibfield  {title} {\bibinfo {title} {Superconducting phases of $f$-electron compounds},\ }\href {https://doi.org/10.1103/RevModPhys.81.1551} {\bibfield  {journal} {\bibinfo  {journal} {Reviews of Modern Physics}\ }\textbf {\bibinfo {volume} {81}},\ \bibinfo {pages} {1551} (\bibinfo {year} {2009})}\BibitemShut {NoStop}%
\bibitem [{\citenamefont {Kuramoto}\ \emph {et~al.}(2009)\citenamefont {Kuramoto}, \citenamefont {Kusunose},\ and\ \citenamefont {Kiss}}]{Kuramoto2009}%
  \BibitemOpen
  \bibfield  {author} {\bibinfo {author} {\bibfnamefont {Y.}~\bibnamefont {Kuramoto}}, \bibinfo {author} {\bibfnamefont {H.}~\bibnamefont {Kusunose}},\ and\ \bibinfo {author} {\bibfnamefont {A.}~\bibnamefont {Kiss}},\ }\bibfield  {title} {\bibinfo {title} {Multipole orders and fluctuations in strongly correlated electron systems},\ }\href {https://doi.org/10.1143/JPSJ.78.072001} {\bibfield  {journal} {\bibinfo  {journal} {Journal of the Physical Society of Japan}\ }\textbf {\bibinfo {volume} {78}},\ \bibinfo {pages} {072001} (\bibinfo {year} {2009})}\BibitemShut {NoStop}%
\bibitem [{\citenamefont {v.~L{\"o}hneysen}\ \emph {et~al.}(2007)\citenamefont {v.~L{\"o}hneysen}, \citenamefont {Rosch}, \citenamefont {Vojta},\ and\ \citenamefont {W{\"o}lfle}}]{Lohneysen2007}%
  \BibitemOpen
  \bibfield  {author} {\bibinfo {author} {\bibfnamefont {H.}~\bibnamefont {v.~L{\"o}hneysen}}, \bibinfo {author} {\bibfnamefont {A.}~\bibnamefont {Rosch}}, \bibinfo {author} {\bibfnamefont {M.}~\bibnamefont {Vojta}},\ and\ \bibinfo {author} {\bibfnamefont {P.}~\bibnamefont {W{\"o}lfle}},\ }\bibfield  {title} {\bibinfo {title} {Fermi-liquid instabilities at magnetic quantum phase transitions},\ }\href {https://doi.org/10.1103/RevModPhys.79.1015} {\bibfield  {journal} {\bibinfo  {journal} {Reviews of Modern Physics}\ }\textbf {\bibinfo {volume} {79}},\ \bibinfo {pages} {1015} (\bibinfo {year} {2007})}\BibitemShut {NoStop}%
\bibitem [{\citenamefont {Mydosh}\ and\ \citenamefont {Oppeneer}(2011)}]{Mydosh2011}%
  \BibitemOpen
  \bibfield  {author} {\bibinfo {author} {\bibfnamefont {J.~A.}\ \bibnamefont {Mydosh}}\ and\ \bibinfo {author} {\bibfnamefont {P.~M.}\ \bibnamefont {Oppeneer}},\ }\bibfield  {title} {\bibinfo {title} {{Colloquium: Hidden order, superconductivity, and magnetism: The unsolved case of URu$_2$Si$_2$}},\ }\href {https://doi.org/10.1103/RevModPhys.83.1301} {\bibfield  {journal} {\bibinfo  {journal} {Rev. Mod. Phys.}\ }\textbf {\bibinfo {volume} {83}},\ \bibinfo {pages} {1301} (\bibinfo {year} {2011})}\BibitemShut {NoStop}%
\bibitem [{\citenamefont {Kondo}(1964)}]{Kondo1964}%
  \BibitemOpen
  \bibfield  {author} {\bibinfo {author} {\bibfnamefont {J.}~\bibnamefont {Kondo}},\ }\bibfield  {title} {\bibinfo {title} {Resistance minimum in dilute magnetic alloys},\ }\href {https://doi.org/10.1143/PTP.32.37} {\bibfield  {journal} {\bibinfo  {journal} {Progress of Theoretical Physics}\ }\textbf {\bibinfo {volume} {32}},\ \bibinfo {pages} {37} (\bibinfo {year} {1964})}\BibitemShut {NoStop}%
\bibitem [{\citenamefont {Ruderman}\ and\ \citenamefont {Kittel}(1954)}]{Ruderman1954}%
  \BibitemOpen
  \bibfield  {author} {\bibinfo {author} {\bibfnamefont {M.~A.}\ \bibnamefont {Ruderman}}\ and\ \bibinfo {author} {\bibfnamefont {C.}~\bibnamefont {Kittel}},\ }\bibfield  {title} {\bibinfo {title} {Indirect exchange coupling of nuclear magnetic moments by conduction electrons},\ }\href {https://doi.org/10.1103/PhysRev.96.99} {\bibfield  {journal} {\bibinfo  {journal} {Phys. Rev.}\ }\textbf {\bibinfo {volume} {96}},\ \bibinfo {pages} {99} (\bibinfo {year} {1954})}\BibitemShut {NoStop}%
\bibitem [{\citenamefont {Kasuya}(1956)}]{Kasuya1956}%
  \BibitemOpen
  \bibfield  {author} {\bibinfo {author} {\bibfnamefont {T.}~\bibnamefont {Kasuya}},\ }\bibfield  {title} {\bibinfo {title} {A theory of metallic ferro- and antiferromagnetism on {Z}ener's model},\ }\href {https://doi.org/10.1143/PTP.16.45} {\bibfield  {journal} {\bibinfo  {journal} {Progress of Theoretical Physics}\ }\textbf {\bibinfo {volume} {16}},\ \bibinfo {pages} {45} (\bibinfo {year} {1956})}\BibitemShut {NoStop}%
\bibitem [{\citenamefont {Yosida}(1957)}]{Yosida1957}%
  \BibitemOpen
  \bibfield  {author} {\bibinfo {author} {\bibfnamefont {K.}~\bibnamefont {Yosida}},\ }\bibfield  {title} {\bibinfo {title} {{Magnetic properties of Cu-Mn alloys}},\ }\href {https://doi.org/10.1103/PhysRev.106.893} {\bibfield  {journal} {\bibinfo  {journal} {Phys. Rev.}\ }\textbf {\bibinfo {volume} {106}},\ \bibinfo {pages} {893} (\bibinfo {year} {1957})}\BibitemShut {NoStop}%
\bibitem [{\citenamefont {Michishita}\ and\ \citenamefont {Peters}(2019)}]{Michishita2019}%
  \BibitemOpen
  \bibfield  {author} {\bibinfo {author} {\bibfnamefont {Y.}~\bibnamefont {Michishita}}\ and\ \bibinfo {author} {\bibfnamefont {R.}~\bibnamefont {Peters}},\ }\bibfield  {title} {\bibinfo {title} {Impact of the {R}ashba spin-orbit coupling on $f$-electron materials},\ }\href {https://doi.org/10.1103/PhysRevB.99.155141} {\bibfield  {journal} {\bibinfo  {journal} {Phys. Rev. B}\ }\textbf {\bibinfo {volume} {99}},\ \bibinfo {pages} {155141} (\bibinfo {year} {2019})}\BibitemShut {NoStop}%
\bibitem [{\citenamefont {Dzsaber}\ \emph {et~al.}(2017)\citenamefont {Dzsaber}, \citenamefont {Yan}, \citenamefont {Eguchi}, \citenamefont {Svagera}, \citenamefont {Prokofiev}, \citenamefont {Chen}, \citenamefont {Tokiwa}, \citenamefont {Gegenwart}, \citenamefont {Kimura}, \citenamefont {Sichelschmidt}, \citenamefont {Coleman},\ and\ \citenamefont {Paschen}}]{Dzsaber2017}%
  \BibitemOpen
  \bibfield  {author} {\bibinfo {author} {\bibfnamefont {S.}~\bibnamefont {Dzsaber}}, \bibinfo {author} {\bibfnamefont {X.}~\bibnamefont {Yan}}, \bibinfo {author} {\bibfnamefont {G.}~\bibnamefont {Eguchi}}, \bibinfo {author} {\bibfnamefont {R.}~\bibnamefont {Svagera}}, \bibinfo {author} {\bibfnamefont {A.}~\bibnamefont {Prokofiev}}, \bibinfo {author} {\bibfnamefont {C.}~\bibnamefont {Chen}}, \bibinfo {author} {\bibfnamefont {Y.}~\bibnamefont {Tokiwa}}, \bibinfo {author} {\bibfnamefont {P.}~\bibnamefont {Gegenwart}}, \bibinfo {author} {\bibfnamefont {S.}~\bibnamefont {Kimura}}, \bibinfo {author} {\bibfnamefont {J.}~\bibnamefont {Sichelschmidt}}, \bibinfo {author} {\bibfnamefont {P.}~\bibnamefont {Coleman}},\ and\ \bibinfo {author} {\bibfnamefont {S.}~\bibnamefont {Paschen}},\ }\bibfield  {title} {\bibinfo {title} {Kondo insulator to semimetal transformation tuned by spin-orbit coupling},\ }\href {https://doi.org/10.1103/PhysRevLett.118.246601} {\bibfield  {journal} {\bibinfo  {journal} {Physical Review
  Letters}\ }\textbf {\bibinfo {volume} {118}},\ \bibinfo {pages} {246601} (\bibinfo {year} {2017})}\BibitemShut {NoStop}%
\bibitem [{\citenamefont {Xu}\ \emph {et~al.}(2014)\citenamefont {Xu}, \citenamefont {Biswas}, \citenamefont {Dil}, \citenamefont {Dhaka}, \citenamefont {Landolt}, \citenamefont {Muff}, \citenamefont {Matt}, \citenamefont {Shi}, \citenamefont {Plumb}, \citenamefont {Radovic}, \citenamefont {Pomjakushina}, \citenamefont {Conder}, \citenamefont {Amato}, \citenamefont {Borisenko}, \citenamefont {Yu}, \citenamefont {Weng}, \citenamefont {Fang}, \citenamefont {Dai}, \citenamefont {Mesot}, \citenamefont {Ding},\ and\ \citenamefont {Radovic}}]{Xu2013}%
  \BibitemOpen
  \bibfield  {author} {\bibinfo {author} {\bibfnamefont {N.}~\bibnamefont {Xu}}, \bibinfo {author} {\bibfnamefont {P.~K.}\ \bibnamefont {Biswas}}, \bibinfo {author} {\bibfnamefont {J.~H.}\ \bibnamefont {Dil}}, \bibinfo {author} {\bibfnamefont {R.~S.}\ \bibnamefont {Dhaka}}, \bibinfo {author} {\bibfnamefont {G.}~\bibnamefont {Landolt}}, \bibinfo {author} {\bibfnamefont {S.}~\bibnamefont {Muff}}, \bibinfo {author} {\bibfnamefont {C.~E.}\ \bibnamefont {Matt}}, \bibinfo {author} {\bibfnamefont {M.}~\bibnamefont {Shi}}, \bibinfo {author} {\bibfnamefont {N.~C.}\ \bibnamefont {Plumb}}, \bibinfo {author} {\bibfnamefont {M.}~\bibnamefont {Radovic}}, \bibinfo {author} {\bibfnamefont {E.}~\bibnamefont {Pomjakushina}}, \bibinfo {author} {\bibfnamefont {K.}~\bibnamefont {Conder}}, \bibinfo {author} {\bibfnamefont {A.}~\bibnamefont {Amato}}, \bibinfo {author} {\bibfnamefont {S.~V.}\ \bibnamefont {Borisenko}}, \bibinfo {author} {\bibfnamefont {R.}~\bibnamefont {Yu}}, \bibinfo {author} {\bibfnamefont {H.~M.}\ \bibnamefont
  {Weng}}, \bibinfo {author} {\bibfnamefont {Z.}~\bibnamefont {Fang}}, \bibinfo {author} {\bibfnamefont {X.}~\bibnamefont {Dai}}, \bibinfo {author} {\bibfnamefont {J.}~\bibnamefont {Mesot}}, \bibinfo {author} {\bibfnamefont {H.}~\bibnamefont {Ding}},\ and\ \bibinfo {author} {\bibfnamefont {M.}~\bibnamefont {Radovic}},\ }\bibfield  {title} {\bibinfo {title} {Surface and bulk electronic structure of the strongly correlated system {SmB}$_6$ and implications for a topological {K}ondo insulator},\ }\href {https://doi.org/10.1038/ncomms5566} {\bibfield  {journal} {\bibinfo  {journal} {Nature Communications}\ }\textbf {\bibinfo {volume} {5}},\ \bibinfo {pages} {4566} (\bibinfo {year} {2014})}\BibitemShut {NoStop}%
\bibitem [{\citenamefont {Ohtsubo}\ \emph {et~al.}(2022)\citenamefont {Ohtsubo}, \citenamefont {Nakaya}, \citenamefont {Nakamura}, \citenamefont {Le~F{\`e}vre}, \citenamefont {Bertran}, \citenamefont {Iga},\ and\ \citenamefont {Kimura}}]{Ohtsubo2022}%
  \BibitemOpen
  \bibfield  {author} {\bibinfo {author} {\bibfnamefont {Y.}~\bibnamefont {Ohtsubo}}, \bibinfo {author} {\bibfnamefont {T.}~\bibnamefont {Nakaya}}, \bibinfo {author} {\bibfnamefont {T.}~\bibnamefont {Nakamura}}, \bibinfo {author} {\bibfnamefont {P.}~\bibnamefont {Le~F{\`e}vre}}, \bibinfo {author} {\bibfnamefont {F.}~\bibnamefont {Bertran}}, \bibinfo {author} {\bibfnamefont {F.}~\bibnamefont {Iga}},\ and\ \bibinfo {author} {\bibfnamefont {S.-I.}\ \bibnamefont {Kimura}},\ }\bibfield  {title} {\bibinfo {title} {Breakdown of bulk-projected isotropy in surface electronic states of topological kondo insulator {SmB}$_6$(001)},\ }\href {https://doi.org/10.1038/s41467-022-33347-0} {\bibfield  {journal} {\bibinfo  {journal} {Nature Communications}\ }\textbf {\bibinfo {volume} {13}},\ \bibinfo {pages} {5600} (\bibinfo {year} {2022})}\BibitemShut {NoStop}%
\bibitem [{\citenamefont {Fumega}\ and\ \citenamefont {Lado}(2024)}]{Fumega2024}%
  \BibitemOpen
  \bibfield  {author} {\bibinfo {author} {\bibfnamefont {A.~O.}\ \bibnamefont {Fumega}}\ and\ \bibinfo {author} {\bibfnamefont {J.~L.}\ \bibnamefont {Lado}},\ }\bibfield  {title} {\bibinfo {title} {Nature of the unconventional heavy-fermion {K}ondo state in monolayer {CeSiI}},\ }\href {https://doi.org/10.1021/acs.nanolett.4c00619} {\bibfield  {journal} {\bibinfo  {journal} {Nano Letters}\ }\textbf {\bibinfo {volume} {24}},\ \bibinfo {pages} {4272} (\bibinfo {year} {2024})}\BibitemShut {NoStop}%
\bibitem [{\citenamefont {Generalov}\ \emph {et~al.}(2018)\citenamefont {Generalov}, \citenamefont {Falke}, \citenamefont {Nechaev}, \citenamefont {Otrokov}, \citenamefont {G\"uttler}, \citenamefont {Chikina}, \citenamefont {Kliemt}, \citenamefont {Seiro}, \citenamefont {Kummer}, \citenamefont {Danzenb\"acher}, \citenamefont {Usachov}, \citenamefont {Kim}, \citenamefont {Dudin}, \citenamefont {Chulkov}, \citenamefont {Laubschat}, \citenamefont {Geibel}, \citenamefont {Krellner},\ and\ \citenamefont {Vyalikh}}]{Generalov2018}%
  \BibitemOpen
  \bibfield  {author} {\bibinfo {author} {\bibfnamefont {A.}~\bibnamefont {Generalov}}, \bibinfo {author} {\bibfnamefont {J.}~\bibnamefont {Falke}}, \bibinfo {author} {\bibfnamefont {I.~A.}\ \bibnamefont {Nechaev}}, \bibinfo {author} {\bibfnamefont {M.~M.}\ \bibnamefont {Otrokov}}, \bibinfo {author} {\bibfnamefont {M.}~\bibnamefont {G\"uttler}}, \bibinfo {author} {\bibfnamefont {A.}~\bibnamefont {Chikina}}, \bibinfo {author} {\bibfnamefont {K.}~\bibnamefont {Kliemt}}, \bibinfo {author} {\bibfnamefont {S.}~\bibnamefont {Seiro}}, \bibinfo {author} {\bibfnamefont {K.}~\bibnamefont {Kummer}}, \bibinfo {author} {\bibfnamefont {S.}~\bibnamefont {Danzenb\"acher}}, \bibinfo {author} {\bibfnamefont {D.}~\bibnamefont {Usachov}}, \bibinfo {author} {\bibfnamefont {T.~K.}\ \bibnamefont {Kim}}, \bibinfo {author} {\bibfnamefont {P.}~\bibnamefont {Dudin}}, \bibinfo {author} {\bibfnamefont {E.~V.}\ \bibnamefont {Chulkov}}, \bibinfo {author} {\bibfnamefont {C.}~\bibnamefont {Laubschat}}, \bibinfo {author} {\bibfnamefont
  {C.}~\bibnamefont {Geibel}}, \bibinfo {author} {\bibfnamefont {C.}~\bibnamefont {Krellner}},\ and\ \bibinfo {author} {\bibfnamefont {D.~V.}\ \bibnamefont {Vyalikh}},\ }\bibfield  {title} {\bibinfo {title} {Strong spin-orbit coupling in the noncentrosymmetric {K}ondo lattice},\ }\href {https://doi.org/10.1103/PhysRevB.98.115157} {\bibfield  {journal} {\bibinfo  {journal} {Phys. Rev. B}\ }\textbf {\bibinfo {volume} {98}},\ \bibinfo {pages} {115157} (\bibinfo {year} {2018})}\BibitemShut {NoStop}%
\bibitem [{\citenamefont {Mende}\ \emph {et~al.}(2022)\citenamefont {Mende}, \citenamefont {Ali}, \citenamefont {Poelchen}, \citenamefont {Schulz}, \citenamefont {Mandic}, \citenamefont {Tarasov}, \citenamefont {Polley}, \citenamefont {Generalov}, \citenamefont {Fedorov}, \citenamefont {Güttler}, \citenamefont {Laubschat}, \citenamefont {Kliemt}, \citenamefont {Koroteev}, \citenamefont {Chulkov}, \citenamefont {Kummer}, \citenamefont {Krellner},\ and\ \citenamefont {Vyalikh}}]{Mende2022}%
  \BibitemOpen
  \bibfield  {author} {\bibinfo {author} {\bibfnamefont {M.}~\bibnamefont {Mende}}, \bibinfo {author} {\bibfnamefont {K.}~\bibnamefont {Ali}}, \bibinfo {author} {\bibfnamefont {G.}~\bibnamefont {Poelchen}}, \bibinfo {author} {\bibfnamefont {S.}~\bibnamefont {Schulz}}, \bibinfo {author} {\bibfnamefont {V.}~\bibnamefont {Mandic}}, \bibinfo {author} {\bibfnamefont {A.~V.}\ \bibnamefont {Tarasov}}, \bibinfo {author} {\bibfnamefont {C.}~\bibnamefont {Polley}}, \bibinfo {author} {\bibfnamefont {A.}~\bibnamefont {Generalov}}, \bibinfo {author} {\bibfnamefont {A.~V.}\ \bibnamefont {Fedorov}}, \bibinfo {author} {\bibfnamefont {M.}~\bibnamefont {Güttler}}, \bibinfo {author} {\bibfnamefont {C.}~\bibnamefont {Laubschat}}, \bibinfo {author} {\bibfnamefont {K.}~\bibnamefont {Kliemt}}, \bibinfo {author} {\bibfnamefont {Y.~M.}\ \bibnamefont {Koroteev}}, \bibinfo {author} {\bibfnamefont {E.~V.}\ \bibnamefont {Chulkov}}, \bibinfo {author} {\bibfnamefont {K.}~\bibnamefont {Kummer}}, \bibinfo {author} {\bibfnamefont
  {C.}~\bibnamefont {Krellner}},\ and\ \bibinfo {author} {\bibfnamefont {D.~V.}\ \bibnamefont {Vyalikh}},\ }\bibfield  {title} {\bibinfo {title} {Strong {R}ashba effect and different $f$-$d$ hybridization phenomena at the surface of the heavy-fermion superconductor {CeIrIn}$_5$},\ }\href {https://doi.org/https://doi.org/10.1002/aelm.202100768} {\bibfield  {journal} {\bibinfo  {journal} {Advanced Electronic Materials}\ }\textbf {\bibinfo {volume} {8}},\ \bibinfo {pages} {2100768} (\bibinfo {year} {2022})}\BibitemShut {NoStop}%
\bibitem [{\citenamefont {Tarasov}\ \emph {et~al.}(2022)\citenamefont {Tarasov}, \citenamefont {Mende}, \citenamefont {Ali}, \citenamefont {Poelchen}, \citenamefont {Schulz}, \citenamefont {Vilkov}, \citenamefont {Bokai}, \citenamefont {Muntwiler}, \citenamefont {Mandic}, \citenamefont {Laubschat}, \citenamefont {Kliemt}, \citenamefont {Krellner}, \citenamefont {Vyalikh},\ and\ \citenamefont {Usachov}}]{Tarasov2022}%
  \BibitemOpen
  \bibfield  {author} {\bibinfo {author} {\bibfnamefont {A.~V.}\ \bibnamefont {Tarasov}}, \bibinfo {author} {\bibfnamefont {M.}~\bibnamefont {Mende}}, \bibinfo {author} {\bibfnamefont {K.}~\bibnamefont {Ali}}, \bibinfo {author} {\bibfnamefont {G.}~\bibnamefont {Poelchen}}, \bibinfo {author} {\bibfnamefont {S.}~\bibnamefont {Schulz}}, \bibinfo {author} {\bibfnamefont {O.~Y.}\ \bibnamefont {Vilkov}}, \bibinfo {author} {\bibfnamefont {K.~A.}\ \bibnamefont {Bokai}}, \bibinfo {author} {\bibfnamefont {M.}~\bibnamefont {Muntwiler}}, \bibinfo {author} {\bibfnamefont {V.}~\bibnamefont {Mandic}}, \bibinfo {author} {\bibfnamefont {C.}~\bibnamefont {Laubschat}}, \bibinfo {author} {\bibfnamefont {K.}~\bibnamefont {Kliemt}}, \bibinfo {author} {\bibfnamefont {C.}~\bibnamefont {Krellner}}, \bibinfo {author} {\bibfnamefont {D.~V.}\ \bibnamefont {Vyalikh}},\ and\ \bibinfo {author} {\bibfnamefont {D.~Y.}\ \bibnamefont {Usachov}},\ }\bibfield  {title} {\bibinfo {title} {Structural instability at the in-terminated surface of the
  heavy-fermion superconductor {CeIrIn}$_5$},\ }\href {https://doi.org/https://doi.org/10.1016/j.surfin.2022.102126} {\bibfield  {journal} {\bibinfo  {journal} {Surfaces and Interfaces}\ }\textbf {\bibinfo {volume} {32}},\ \bibinfo {pages} {102126} (\bibinfo {year} {2022})}\BibitemShut {NoStop}%
\bibitem [{\citenamefont {Lai}\ \emph {et~al.}(2018)\citenamefont {Lai}, \citenamefont {Grefe}, \citenamefont {Paschen},\ and\ \citenamefont {Si}}]{Lai2018}%
  \BibitemOpen
  \bibfield  {author} {\bibinfo {author} {\bibfnamefont {H.-H.}\ \bibnamefont {Lai}}, \bibinfo {author} {\bibfnamefont {S.~E.}\ \bibnamefont {Grefe}}, \bibinfo {author} {\bibfnamefont {S.}~\bibnamefont {Paschen}},\ and\ \bibinfo {author} {\bibfnamefont {Q.}~\bibnamefont {Si}},\ }\bibfield  {title} {\bibinfo {title} {Weyl–{K}ondo semimetal in heavy-fermion systems},\ }\href {https://doi.org/10.1073/pnas.1715851115} {\bibfield  {journal} {\bibinfo  {journal} {Proceedings of the National Academy of Sciences}\ }\textbf {\bibinfo {volume} {115}},\ \bibinfo {pages} {93} (\bibinfo {year} {2018})}\BibitemShut {NoStop}%
\bibitem [{\citenamefont {Grefe}\ \emph {et~al.}(2020)\citenamefont {Grefe}, \citenamefont {Lai}, \citenamefont {Paschen},\ and\ \citenamefont {Si}}]{Grefe2020}%
  \BibitemOpen
  \bibfield  {author} {\bibinfo {author} {\bibfnamefont {S.~E.}\ \bibnamefont {Grefe}}, \bibinfo {author} {\bibfnamefont {H.-H.}\ \bibnamefont {Lai}}, \bibinfo {author} {\bibfnamefont {S.}~\bibnamefont {Paschen}},\ and\ \bibinfo {author} {\bibfnamefont {Q.}~\bibnamefont {Si}},\ }\bibfield  {title} {\bibinfo {title} {Weyl-{K}ondo semimetals in nonsymmorphic systems},\ }\href {https://doi.org/10.1103/PhysRevB.101.075138} {\bibfield  {journal} {\bibinfo  {journal} {Phys. Rev. B}\ }\textbf {\bibinfo {volume} {101}},\ \bibinfo {pages} {075138} (\bibinfo {year} {2020})}\BibitemShut {NoStop}%
\bibitem [{\citenamefont {Dzsaber}\ \emph {et~al.}(2021)\citenamefont {Dzsaber}, \citenamefont {Yan}, \citenamefont {Taupin}, \citenamefont {Eguchi}, \citenamefont {Prokofiev}, \citenamefont {Shiroka}, \citenamefont {Blaha}, \citenamefont {Rubel}, \citenamefont {Grefe}, \citenamefont {Lai}, \citenamefont {Si},\ and\ \citenamefont {Paschen}}]{Samir2021}%
  \BibitemOpen
  \bibfield  {author} {\bibinfo {author} {\bibfnamefont {S.}~\bibnamefont {Dzsaber}}, \bibinfo {author} {\bibfnamefont {X.}~\bibnamefont {Yan}}, \bibinfo {author} {\bibfnamefont {M.}~\bibnamefont {Taupin}}, \bibinfo {author} {\bibfnamefont {G.}~\bibnamefont {Eguchi}}, \bibinfo {author} {\bibfnamefont {A.}~\bibnamefont {Prokofiev}}, \bibinfo {author} {\bibfnamefont {T.}~\bibnamefont {Shiroka}}, \bibinfo {author} {\bibfnamefont {P.}~\bibnamefont {Blaha}}, \bibinfo {author} {\bibfnamefont {O.}~\bibnamefont {Rubel}}, \bibinfo {author} {\bibfnamefont {S.~E.}\ \bibnamefont {Grefe}}, \bibinfo {author} {\bibfnamefont {H.-H.}\ \bibnamefont {Lai}}, \bibinfo {author} {\bibfnamefont {Q.}~\bibnamefont {Si}},\ and\ \bibinfo {author} {\bibfnamefont {S.}~\bibnamefont {Paschen}},\ }\bibfield  {title} {\bibinfo {title} {Giant spontaneous {H}all effect in a nonmagnetic {Weyl–Kondo} semimetal},\ }\href {https://doi.org/10.1073/pnas.2013386118} {\bibfield  {journal} {\bibinfo  {journal} {Proceedings of the National Academy of
  Sciences}\ }\textbf {\bibinfo {volume} {118}},\ \bibinfo {pages} {e2013386118} (\bibinfo {year} {2021})}\BibitemShut {NoStop}%
\bibitem [{\citenamefont {Neupane}\ \emph {et~al.}(2013)\citenamefont {Neupane}, \citenamefont {Alidoust}, \citenamefont {Xu}, \citenamefont {Kondo}, \citenamefont {Ishida}, \citenamefont {Kim}, \citenamefont {Liu}, \citenamefont {Belopolski}, \citenamefont {Jo}, \citenamefont {Chang}, \citenamefont {Jeng}, \citenamefont {Durakiewicz}, \citenamefont {Balatsky}, \citenamefont {Shin},\ and\ \citenamefont {Zahid~Hasan}}]{Neupane2013}%
  \BibitemOpen
  \bibfield  {author} {\bibinfo {author} {\bibfnamefont {M.}~\bibnamefont {Neupane}}, \bibinfo {author} {\bibfnamefont {N.}~\bibnamefont {Alidoust}}, \bibinfo {author} {\bibfnamefont {S.-Y.}\ \bibnamefont {Xu}}, \bibinfo {author} {\bibfnamefont {T.}~\bibnamefont {Kondo}}, \bibinfo {author} {\bibfnamefont {Y.}~\bibnamefont {Ishida}}, \bibinfo {author} {\bibfnamefont {D.~J.}\ \bibnamefont {Kim}}, \bibinfo {author} {\bibfnamefont {C.}~\bibnamefont {Liu}}, \bibinfo {author} {\bibfnamefont {I.}~\bibnamefont {Belopolski}}, \bibinfo {author} {\bibfnamefont {N.~H.}\ \bibnamefont {Jo}}, \bibinfo {author} {\bibfnamefont {T.-R.}\ \bibnamefont {Chang}}, \bibinfo {author} {\bibfnamefont {H.}~\bibnamefont {Jeng}}, \bibinfo {author} {\bibfnamefont {T.}~\bibnamefont {Durakiewicz}}, \bibinfo {author} {\bibfnamefont {A.~V.}\ \bibnamefont {Balatsky}}, \bibinfo {author} {\bibfnamefont {S.}~\bibnamefont {Shin}},\ and\ \bibinfo {author} {\bibfnamefont {M.}~\bibnamefont {Zahid~Hasan}},\ }\bibfield  {title} {\bibinfo {title}
  {Surface electronic structure of the topological {K}ondo-insulator candidate correlated electron system {SmB}$_6$},\ }\href {https://doi.org/10.1038/ncomms3991} {\bibfield  {journal} {\bibinfo  {journal} {Nature Communications}\ }\textbf {\bibinfo {volume} {4}},\ \bibinfo {pages} {2991} (\bibinfo {year} {2013})}\BibitemShut {NoStop}%
\bibitem [{\citenamefont {Weng}\ \emph {et~al.}(2014)\citenamefont {Weng}, \citenamefont {Zhao}, \citenamefont {Wang}, \citenamefont {Fang}, \citenamefont {Dai},\ and\ \citenamefont {Shi}}]{Weng2014}%
  \BibitemOpen
  \bibfield  {author} {\bibinfo {author} {\bibfnamefont {H.}~\bibnamefont {Weng}}, \bibinfo {author} {\bibfnamefont {J.}~\bibnamefont {Zhao}}, \bibinfo {author} {\bibfnamefont {Z.~F.}\ \bibnamefont {Wang}}, \bibinfo {author} {\bibfnamefont {Z.}~\bibnamefont {Fang}}, \bibinfo {author} {\bibfnamefont {X.}~\bibnamefont {Dai}},\ and\ \bibinfo {author} {\bibfnamefont {Y.}~\bibnamefont {Shi}},\ }\bibfield  {title} {\bibinfo {title} {Topological crystalline {K}ondo insulator in mixed valence {Y}tterbium {B}orides},\ }\href {https://doi.org/10.1103/PhysRevLett.112.016403} {\bibfield  {journal} {\bibinfo  {journal} {Physical Review Letters}\ }\textbf {\bibinfo {volume} {112}},\ \bibinfo {pages} {016403} (\bibinfo {year} {2014})}\BibitemShut {NoStop}%
\bibitem [{\citenamefont {Dzero}\ \emph {et~al.}(2016)\citenamefont {Dzero}, \citenamefont {Xia}, \citenamefont {Galitski},\ and\ \citenamefont {Coleman}}]{Dzero2016}%
  \BibitemOpen
  \bibfield  {author} {\bibinfo {author} {\bibfnamefont {M.}~\bibnamefont {Dzero}}, \bibinfo {author} {\bibfnamefont {J.}~\bibnamefont {Xia}}, \bibinfo {author} {\bibfnamefont {V.}~\bibnamefont {Galitski}},\ and\ \bibinfo {author} {\bibfnamefont {P.}~\bibnamefont {Coleman}},\ }\bibfield  {title} {\bibinfo {title} {Topological {K}ondo insulators},\ }\href {https://doi.org/https://doi.org/10.1146/annurev-conmatphys-031214-014749} {\bibfield  {journal} {\bibinfo  {journal} {Annual Review of Condensed Matter Physics}\ }\textbf {\bibinfo {volume} {7}},\ \bibinfo {pages} {249} (\bibinfo {year} {2016})}\BibitemShut {NoStop}%
\bibitem [{\citenamefont {Vidhyadhiraja}\ and\ \citenamefont {Logan}(2004)}]{Vidhyadhiraja2004}%
  \BibitemOpen
  \bibfield  {author} {\bibinfo {author} {\bibfnamefont {N.~S.}\ \bibnamefont {Vidhyadhiraja}}\ and\ \bibinfo {author} {\bibfnamefont {D.~E.}\ \bibnamefont {Logan}},\ }\bibfield  {title} {\bibinfo {title} {{Dynamics and scaling in the periodic Anderson model}},\ }\href {https://doi.org/10.1140/epjb/e2004-00197-6} {\bibfield  {journal} {\bibinfo  {journal} {Eur. Phys. J. B}\ }\textbf {\bibinfo {volume} {39}},\ \bibinfo {pages} {313} (\bibinfo {year} {2004})},\ \Eprint {https://arxiv.org/abs/cond-mat/0406009} {cond-mat/0406009} \BibitemShut {NoStop}%
\bibitem [{\citenamefont {Hagym\'asi}\ \emph {et~al.}(2012)\citenamefont {Hagym\'asi}, \citenamefont {Itai},\ and\ \citenamefont {S\'olyom}}]{Hagymasi2011}%
  \BibitemOpen
  \bibfield  {author} {\bibinfo {author} {\bibfnamefont {I.}~\bibnamefont {Hagym\'asi}}, \bibinfo {author} {\bibfnamefont {K.}~\bibnamefont {Itai}},\ and\ \bibinfo {author} {\bibfnamefont {J.}~\bibnamefont {S\'olyom}},\ }\bibfield  {title} {\bibinfo {title} {Periodic {A}nderson model with correlated conduction electrons: Variational and exact diagonalization study},\ }\href {https://doi.org/10.1103/PhysRevB.85.235116} {\bibfield  {journal} {\bibinfo  {journal} {Phys. Rev. B}\ }\textbf {\bibinfo {volume} {85}},\ \bibinfo {pages} {235116} (\bibinfo {year} {2012})}\BibitemShut {NoStop}%
\bibitem [{\citenamefont {Costa}\ \emph {et~al.}(2018)\citenamefont {Costa}, \citenamefont {Ara\'ujo}, \citenamefont {Lima}, \citenamefont {Paiva}, \citenamefont {dos Santos},\ and\ \citenamefont {Scalettar}}]{Costa2018}%
  \BibitemOpen
  \bibfield  {author} {\bibinfo {author} {\bibfnamefont {N.~C.}\ \bibnamefont {Costa}}, \bibinfo {author} {\bibfnamefont {M.~V.}\ \bibnamefont {Ara\'ujo}}, \bibinfo {author} {\bibfnamefont {J.~P.}\ \bibnamefont {Lima}}, \bibinfo {author} {\bibfnamefont {T.}~\bibnamefont {Paiva}}, \bibinfo {author} {\bibfnamefont {R.~R.}\ \bibnamefont {dos Santos}},\ and\ \bibinfo {author} {\bibfnamefont {R.~T.}\ \bibnamefont {Scalettar}},\ }\bibfield  {title} {\bibinfo {title} {{Compressible ferrimagnetism in the depleted periodic Anderson model}},\ }\href {https://doi.org/10.1103/PhysRevB.97.085123} {\bibfield  {journal} {\bibinfo  {journal} {Phys. Rev. B}\ }\textbf {\bibinfo {volume} {97}},\ \bibinfo {pages} {085123} (\bibinfo {year} {2018})}\BibitemShut {NoStop}%
\bibitem [{\citenamefont {Costa}\ \emph {et~al.}(2019)\citenamefont {Costa}, \citenamefont {Mendes-Santos}, \citenamefont {Paiva}, \citenamefont {Curro}, \citenamefont {dos Santos},\ and\ \citenamefont {Scalettar}}]{Costa2019}%
  \BibitemOpen
  \bibfield  {author} {\bibinfo {author} {\bibfnamefont {N.~C.}\ \bibnamefont {Costa}}, \bibinfo {author} {\bibfnamefont {T.}~\bibnamefont {Mendes-Santos}}, \bibinfo {author} {\bibfnamefont {T.}~\bibnamefont {Paiva}}, \bibinfo {author} {\bibfnamefont {N.~J.}\ \bibnamefont {Curro}}, \bibinfo {author} {\bibfnamefont {R.~R.}\ \bibnamefont {dos Santos}},\ and\ \bibinfo {author} {\bibfnamefont {R.~T.}\ \bibnamefont {Scalettar}},\ }\bibfield  {title} {\bibinfo {title} {Coherence temperature in the diluted periodic {A}nderson model},\ }\href {https://doi.org/10.1103/PhysRevB.99.195116} {\bibfield  {journal} {\bibinfo  {journal} {Phys. Rev. B}\ }\textbf {\bibinfo {volume} {99}},\ \bibinfo {pages} {195116} (\bibinfo {year} {2019})}\BibitemShut {NoStop}%
\bibitem [{\citenamefont {Zhang}\ \emph {et~al.}(2019)\citenamefont {Zhang}, \citenamefont {Ma}, \citenamefont {Costa}, \citenamefont {dos Santos},\ and\ \citenamefont {Scalettar}}]{Zhang2019}%
  \BibitemOpen
  \bibfield  {author} {\bibinfo {author} {\bibfnamefont {L.}~\bibnamefont {Zhang}}, \bibinfo {author} {\bibfnamefont {T.}~\bibnamefont {Ma}}, \bibinfo {author} {\bibfnamefont {N.~C.}\ \bibnamefont {Costa}}, \bibinfo {author} {\bibfnamefont {R.~R.}\ \bibnamefont {dos Santos}},\ and\ \bibinfo {author} {\bibfnamefont {R.~T.}\ \bibnamefont {Scalettar}},\ }\bibfield  {title} {\bibinfo {title} {{Determinant quantum Monte Carlo study of exhaustion in the periodic Anderson model}},\ }\href {https://doi.org/10.1103/PhysRevB.99.195147} {\bibfield  {journal} {\bibinfo  {journal} {Phys. Rev. B}\ }\textbf {\bibinfo {volume} {99}},\ \bibinfo {pages} {195147} (\bibinfo {year} {2019})}\BibitemShut {NoStop}%
\bibitem [{\citenamefont {Oliveira}\ \emph {et~al.}(2023)\citenamefont {Oliveira}, \citenamefont {Paiva}, \citenamefont {Scalettar},\ and\ \citenamefont {Costa}}]{Oliveira2023}%
  \BibitemOpen
  \bibfield  {author} {\bibinfo {author} {\bibfnamefont {W.~S.}\ \bibnamefont {Oliveira}}, \bibinfo {author} {\bibfnamefont {T.}~\bibnamefont {Paiva}}, \bibinfo {author} {\bibfnamefont {R.~T.}\ \bibnamefont {Scalettar}},\ and\ \bibinfo {author} {\bibfnamefont {N.~C.}\ \bibnamefont {Costa}},\ }\bibfield  {title} {\bibinfo {title} {Magnetic and singlet phases in the three-dimensional periodic {A}nderson model},\ }\href {https://doi.org/10.1103/PhysRevB.108.115121} {\bibfield  {journal} {\bibinfo  {journal} {Phys. Rev. B}\ }\textbf {\bibinfo {volume} {108}},\ \bibinfo {pages} {115121} (\bibinfo {year} {2023})}\BibitemShut {NoStop}%
\bibitem [{\citenamefont {Gleis}\ \emph {et~al.}(2024)\citenamefont {Gleis}, \citenamefont {Lee}, \citenamefont {Kotliar},\ and\ \citenamefont {von Delft}}]{Gleis2024}%
  \BibitemOpen
  \bibfield  {author} {\bibinfo {author} {\bibfnamefont {A.}~\bibnamefont {Gleis}}, \bibinfo {author} {\bibfnamefont {S.-S.~B.}\ \bibnamefont {Lee}}, \bibinfo {author} {\bibfnamefont {G.}~\bibnamefont {Kotliar}},\ and\ \bibinfo {author} {\bibfnamefont {J.}~\bibnamefont {von Delft}},\ }\bibfield  {title} {\bibinfo {title} {Emergent properties of the periodic {A}nderson model: A high-resolution, real-frequency study of heavy-fermion quantum criticality},\ }\href {https://doi.org/10.1103/PhysRevX.14.041036} {\bibfield  {journal} {\bibinfo  {journal} {Phys. Rev. X}\ }\textbf {\bibinfo {volume} {14}},\ \bibinfo {pages} {041036} (\bibinfo {year} {2024})}\BibitemShut {NoStop}%
\bibitem [{\citenamefont {Hagym\'asi}(2025)}]{Hagymasi2025}%
  \BibitemOpen
  \bibfield  {author} {\bibinfo {author} {\bibfnamefont {I.}~\bibnamefont {Hagym\'asi}},\ }\bibfield  {title} {\bibinfo {title} {Magnetic phases of the periodic {A}nderson model in two dimensions},\ }\href {https://doi.org/10.1103/dh8r-v24r} {\bibfield  {journal} {\bibinfo  {journal} {Phys. Rev. B}\ }\textbf {\bibinfo {volume} {112}},\ \bibinfo {pages} {045118} (\bibinfo {year} {2025})}\BibitemShut {NoStop}%
\bibitem [{\citenamefont {Tanaka}\ and\ \citenamefont {Kontani}(2010)}]{Tanaka2010}%
  \BibitemOpen
  \bibfield  {author} {\bibinfo {author} {\bibfnamefont {T.}~\bibnamefont {Tanaka}}\ and\ \bibinfo {author} {\bibfnamefont {H.}~\bibnamefont {Kontani}},\ }\bibfield  {title} {\bibinfo {title} {{Intrinsic spin and orbital Hall effects in heavy-fermion systems}},\ }\href {https://doi.org/10.1103/PhysRevB.81.224401} {\bibfield  {journal} {\bibinfo  {journal} {Phys. Rev. B}\ }\textbf {\bibinfo {volume} {81}},\ \bibinfo {pages} {224401} (\bibinfo {year} {2010})}\BibitemShut {NoStop}%
\bibitem [{\citenamefont {Yambe}\ and\ \citenamefont {Hayami}(2022)}]{Yambe2022}%
  \BibitemOpen
  \bibfield  {author} {\bibinfo {author} {\bibfnamefont {R.}~\bibnamefont {Yambe}}\ and\ \bibinfo {author} {\bibfnamefont {S.}~\bibnamefont {Hayami}},\ }\bibfield  {title} {\bibinfo {title} {{Effective spin model in momentum space: Toward a systematic understanding of multiple-instability by momentum-resolved anisotropic exchange interactions}},\ }\href {https://doi.org/10.1103/PhysRevB.106.174437} {\bibfield  {journal} {\bibinfo  {journal} {Phys. Rev. B}\ }\textbf {\bibinfo {volume} {106}},\ \bibinfo {pages} {174437} (\bibinfo {year} {2022})}\BibitemShut {NoStop}%
\bibitem [{\citenamefont {Harima}(1987)}]{Harima1987}%
  \BibitemOpen
  \bibfield  {author} {\bibinfo {author} {\bibfnamefont {H.}~\bibnamefont {Harima}},\ }\bibfield  {title} {\bibinfo {title} {{Susceptibility of the Periodic Anderson Model with Spin–Orbit Coupling}},\ }\href {https://doi.org/10.1143/PTP.77.1116} {\bibfield  {journal} {\bibinfo  {journal} {Prog. Theor. Phys.}\ }\textbf {\bibinfo {volume} {77}},\ \bibinfo {pages} {1116} (\bibinfo {year} {1987})}\BibitemShut {NoStop}%
\bibitem [{\citenamefont {Dzero}\ \emph {et~al.}(2010)\citenamefont {Dzero}, \citenamefont {Sun}, \citenamefont {Galitski},\ and\ \citenamefont {Coleman}}]{Dzero2010}%
  \BibitemOpen
  \bibfield  {author} {\bibinfo {author} {\bibfnamefont {M.}~\bibnamefont {Dzero}}, \bibinfo {author} {\bibfnamefont {K.}~\bibnamefont {Sun}}, \bibinfo {author} {\bibfnamefont {V.}~\bibnamefont {Galitski}},\ and\ \bibinfo {author} {\bibfnamefont {P.}~\bibnamefont {Coleman}},\ }\bibfield  {title} {\bibinfo {title} {Topological {K}ondo insulators},\ }\href {https://doi.org/10.1103/PhysRevLett.104.106408} {\bibfield  {journal} {\bibinfo  {journal} {Phys. Rev. Lett.}\ }\textbf {\bibinfo {volume} {104}},\ \bibinfo {pages} {106408} (\bibinfo {year} {2010})}\BibitemShut {NoStop}%
\bibitem [{\citenamefont {Dzero}\ \emph {et~al.}(2012)\citenamefont {Dzero}, \citenamefont {Sun}, \citenamefont {Coleman},\ and\ \citenamefont {Galitski}}]{Dzero2012}%
  \BibitemOpen
  \bibfield  {author} {\bibinfo {author} {\bibfnamefont {M.}~\bibnamefont {Dzero}}, \bibinfo {author} {\bibfnamefont {K.}~\bibnamefont {Sun}}, \bibinfo {author} {\bibfnamefont {P.}~\bibnamefont {Coleman}},\ and\ \bibinfo {author} {\bibfnamefont {V.}~\bibnamefont {Galitski}},\ }\bibfield  {title} {\bibinfo {title} {Theory of topological {K}ondo insulators},\ }\href {https://doi.org/10.1103/PhysRevB.85.045130} {\bibfield  {journal} {\bibinfo  {journal} {Phys. Rev. B}\ }\textbf {\bibinfo {volume} {85}},\ \bibinfo {pages} {045130} (\bibinfo {year} {2012})}\BibitemShut {NoStop}%
\bibitem [{\citenamefont {Legner}\ \emph {et~al.}(2014)\citenamefont {Legner}, \citenamefont {R\"uegg},\ and\ \citenamefont {Sigrist}}]{Legner2014}%
  \BibitemOpen
  \bibfield  {author} {\bibinfo {author} {\bibfnamefont {M.}~\bibnamefont {Legner}}, \bibinfo {author} {\bibfnamefont {A.}~\bibnamefont {R\"uegg}},\ and\ \bibinfo {author} {\bibfnamefont {M.}~\bibnamefont {Sigrist}},\ }\bibfield  {title} {\bibinfo {title} {Topological invariants, surface states, and interaction-driven phase transitions in correlated {K}ondo insulators with cubic symmetry},\ }\href {https://doi.org/10.1103/PhysRevB.89.085110} {\bibfield  {journal} {\bibinfo  {journal} {Phys. Rev. B}\ }\textbf {\bibinfo {volume} {89}},\ \bibinfo {pages} {085110} (\bibinfo {year} {2014})}\BibitemShut {NoStop}%
\bibitem [{\citenamefont {Lu}\ \emph {et~al.}(2019)\citenamefont {Lu}, \citenamefont {Chou}, \citenamefont {Chung},\ and\ \citenamefont {Mou}}]{Lu2019}%
  \BibitemOpen
  \bibfield  {author} {\bibinfo {author} {\bibfnamefont {Y.-W.}\ \bibnamefont {Lu}}, \bibinfo {author} {\bibfnamefont {P.-H.}\ \bibnamefont {Chou}}, \bibinfo {author} {\bibfnamefont {C.-H.}\ \bibnamefont {Chung}},\ and\ \bibinfo {author} {\bibfnamefont {C.-Y.}\ \bibnamefont {Mou}},\ }\bibfield  {title} {\bibinfo {title} {Tunable topological semimetallic phases in {K}ondo lattice systems},\ }\href {https://doi.org/10.1103/PhysRevB.99.035141} {\bibfield  {journal} {\bibinfo  {journal} {Phys. Rev. B}\ }\textbf {\bibinfo {volume} {99}},\ \bibinfo {pages} {035141} (\bibinfo {year} {2019})}\BibitemShut {NoStop}%
\bibitem [{\citenamefont {Tang}\ \emph {et~al.}(2014)\citenamefont {Tang}, \citenamefont {Yang}, \citenamefont {Sun},\ and\ \citenamefont {Lin}}]{Tang2014}%
  \BibitemOpen
  \bibfield  {author} {\bibinfo {author} {\bibfnamefont {H.-K.}\ \bibnamefont {Tang}}, \bibinfo {author} {\bibfnamefont {X.}~\bibnamefont {Yang}}, \bibinfo {author} {\bibfnamefont {J.}~\bibnamefont {Sun}},\ and\ \bibinfo {author} {\bibfnamefont {H.-Q.}\ \bibnamefont {Lin}},\ }\bibfield  {title} {\bibinfo {title} {{Berezinskii-Kosterlitz-Thouless phase transition of spin-orbit coupled Fermi gas in optical lattice}},\ }\href {https://doi.org/10.1209/0295-5075/107/40003} {\bibfield  {journal} {\bibinfo  {journal} {Europhysics Letters}\ }\textbf {\bibinfo {volume} {107}},\ \bibinfo {pages} {40003} (\bibinfo {year} {2014})}\BibitemShut {NoStop}%
\bibitem [{\citenamefont {Rosenberg}\ \emph {et~al.}(2017)\citenamefont {Rosenberg}, \citenamefont {Shi},\ and\ \citenamefont {Zhang}}]{Rosenberg2017}%
  \BibitemOpen
  \bibfield  {author} {\bibinfo {author} {\bibfnamefont {P.}~\bibnamefont {Rosenberg}}, \bibinfo {author} {\bibfnamefont {H.}~\bibnamefont {Shi}},\ and\ \bibinfo {author} {\bibfnamefont {S.}~\bibnamefont {Zhang}},\ }\bibfield  {title} {\bibinfo {title} {{Ultracold Atoms in a Square Lattice with Spin-Orbit Coupling: Charge Order, Superfluidity, and Topological Signatures}},\ }\href {https://doi.org/10.1103/PhysRevLett.119.265301} {\bibfield  {journal} {\bibinfo  {journal} {Phys. Rev. Lett.}\ }\textbf {\bibinfo {volume} {119}},\ \bibinfo {pages} {265301} (\bibinfo {year} {2017})}\BibitemShut {NoStop}%
\bibitem [{\citenamefont {Kim}\ \emph {et~al.}(2020)\citenamefont {Kim}, \citenamefont {Werner},\ and\ \citenamefont {Valenti}}]{Kim2020}%
  \BibitemOpen
  \bibfield  {author} {\bibinfo {author} {\bibfnamefont {A.~J.}\ \bibnamefont {Kim}}, \bibinfo {author} {\bibfnamefont {P.}~\bibnamefont {Werner}},\ and\ \bibinfo {author} {\bibfnamefont {R.}~\bibnamefont {Valenti}},\ }\bibfield  {title} {\bibinfo {title} {Alleviating the sign problem in quantum monte carlo simulations of spin-orbit-coupled multiorbital {H}ubbard models},\ }\href {https://doi.org/10.1103/PhysRevB.101.045108} {\bibfield  {journal} {\bibinfo  {journal} {Phys. Rev. B}\ }\textbf {\bibinfo {volume} {101}},\ \bibinfo {pages} {045108} (\bibinfo {year} {2020})}\BibitemShut {NoStop}%
\bibitem [{\citenamefont {Wan}\ \emph {et~al.}(2022)\citenamefont {Wan}, \citenamefont {Zhang},\ and\ \citenamefont {Yao}}]{Wan2022}%
  \BibitemOpen
  \bibfield  {author} {\bibinfo {author} {\bibfnamefont {Z.-Q.}\ \bibnamefont {Wan}}, \bibinfo {author} {\bibfnamefont {S.-X.}\ \bibnamefont {Zhang}},\ and\ \bibinfo {author} {\bibfnamefont {H.}~\bibnamefont {Yao}},\ }\bibfield  {title} {\bibinfo {title} {{Mitigating the fermion sign problem by automatic differentiation}},\ }\href {https://doi.org/10.1103/PhysRevB.106.L241109} {\bibfield  {journal} {\bibinfo  {journal} {Phys. Rev. B}\ }\textbf {\bibinfo {volume} {106}},\ \bibinfo {pages} {L241109} (\bibinfo {year} {2022})}\BibitemShut {NoStop}%
\bibitem [{\citenamefont {Sousa-Jr.}\ and\ \citenamefont {Mondaini}(2025)}]{SousaJunior2025}%
  \BibitemOpen
  \bibfield  {author} {\bibinfo {author} {\bibfnamefont {S.~a. d.~A.}\ \bibnamefont {Sousa-Jr.}}\ and\ \bibinfo {author} {\bibfnamefont {R.}~\bibnamefont {Mondaini}},\ }\bibfield  {title} {\bibinfo {title} {{Weyl semimetallic, N\'eel, spiral, and vortex states in the Rashba-Hubbard model}},\ }\href {https://doi.org/10.1103/PhysRevB.111.075166} {\bibfield  {journal} {\bibinfo  {journal} {Phys. Rev. B}\ }\textbf {\bibinfo {volume} {111}},\ \bibinfo {pages} {075166} (\bibinfo {year} {2025})}\BibitemShut {NoStop}%
\bibitem [{\citenamefont {Faúndez}\ \emph {et~al.}(2025)\citenamefont {Faúndez}, \citenamefont {Fontenele}, \citenamefont {dos Anjos Sousa-Júnior}, \citenamefont {Assaad},\ and\ \citenamefont {Costa}}]{Faundez2024}%
  \BibitemOpen
  \bibfield  {author} {\bibinfo {author} {\bibfnamefont {J.}~\bibnamefont {Faúndez}}, \bibinfo {author} {\bibfnamefont {R.~A.}\ \bibnamefont {Fontenele}}, \bibinfo {author} {\bibfnamefont {S.}~\bibnamefont {dos Anjos Sousa-Júnior}}, \bibinfo {author} {\bibfnamefont {F.~F.}\ \bibnamefont {Assaad}},\ and\ \bibinfo {author} {\bibfnamefont {N.~C.}\ \bibnamefont {Costa}},\ }\href {https://arxiv.org/abs/2411.07119} {\bibinfo {title} {The two-dimensional {Rashba-Holstein} model: A quantum {Monte Carlo} approach}} (\bibinfo {year} {2025}),\ \Eprint {https://arxiv.org/abs/2411.07119} {arXiv:2411.07119 [cond-mat.str-el]} \BibitemShut {NoStop}%
\bibitem [{\citenamefont {Luo}\ \emph {et~al.}(2021)\citenamefont {Luo}, \citenamefont {Ferrero}, \citenamefont {Yao},\ and\ \citenamefont {Wu}}]{Luo2021}%
  \BibitemOpen
  \bibfield  {author} {\bibinfo {author} {\bibfnamefont {Z.}~\bibnamefont {Luo}}, \bibinfo {author} {\bibfnamefont {M.}~\bibnamefont {Ferrero}}, \bibinfo {author} {\bibfnamefont {D.-X.}\ \bibnamefont {Yao}},\ and\ \bibinfo {author} {\bibfnamefont {W.}~\bibnamefont {Wu}},\ }\bibfield  {title} {\bibinfo {title} {{Inexorable edge Kondo breakdown in topological Kondo insulators}},\ }\href {https://doi.org/10.1103/PhysRevB.104.L161119} {\bibfield  {journal} {\bibinfo  {journal} {Phys. Rev. B}\ }\textbf {\bibinfo {volume} {104}},\ \bibinfo {pages} {L161119} (\bibinfo {year} {2021})}\BibitemShut {NoStop}%
\bibitem [{\citenamefont {Blankenbecler}\ \emph {et~al.}(1981)\citenamefont {Blankenbecler}, \citenamefont {Scalapino},\ and\ \citenamefont {Sugar}}]{Blankenbecler1981}%
  \BibitemOpen
  \bibfield  {author} {\bibinfo {author} {\bibfnamefont {R.}~\bibnamefont {Blankenbecler}}, \bibinfo {author} {\bibfnamefont {D.~J.}\ \bibnamefont {Scalapino}},\ and\ \bibinfo {author} {\bibfnamefont {R.~L.}\ \bibnamefont {Sugar}},\ }\bibfield  {title} {\bibinfo {title} {{Monte Carlo calculations of coupled boson-fermion systems. I}},\ }\href {https://doi.org/10.1103/PhysRevD.24.2278} {\bibfield  {journal} {\bibinfo  {journal} {Phys. Rev. D}\ }\textbf {\bibinfo {volume} {24}},\ \bibinfo {pages} {2278} (\bibinfo {year} {1981})}\BibitemShut {NoStop}%
\bibitem [{\citenamefont {Hirsch}(1985)}]{Hirsch1985}%
  \BibitemOpen
  \bibfield  {author} {\bibinfo {author} {\bibfnamefont {J.~E.}\ \bibnamefont {Hirsch}},\ }\bibfield  {title} {\bibinfo {title} {{Two-dimensional Hubbard model: Numerical simulation study}},\ }\href {https://doi.org/10.1103/PhysRevB.31.4403} {\bibfield  {journal} {\bibinfo  {journal} {Phys. Rev. B}\ }\textbf {\bibinfo {volume} {31}},\ \bibinfo {pages} {4403} (\bibinfo {year} {1985})}\BibitemShut {NoStop}%
\bibitem [{\citenamefont {dos Santos}(2003)}]{rrds2003}%
  \BibitemOpen
  \bibfield  {author} {\bibinfo {author} {\bibfnamefont {R.~R.}\ \bibnamefont {dos Santos}},\ }\bibfield  {title} {\bibinfo {title} {{Introduction to quantum Monte Carlo simulations for fermionic systems}},\ }\href {https://doi.org/10.1590/S0103-97332003000100003} {\bibfield  {journal} {\bibinfo  {journal} {Braz. J. Phys}\ }\textbf {\bibinfo {volume} {33}},\ \bibinfo {pages} {63} (\bibinfo {year} {2003})}\BibitemShut {NoStop}%
\bibitem [{\citenamefont {Assaad}\ and\ \citenamefont {Evertz}(2008)}]{Assaad2008}%
  \BibitemOpen
  \bibfield  {author} {\bibinfo {author} {\bibfnamefont {F.}~\bibnamefont {Assaad}}\ and\ \bibinfo {author} {\bibfnamefont {H.}~\bibnamefont {Evertz}},\ }\bibinfo {title} {{World-line and Determinantal Quantum {M}onte {C}arlo Methods for Spins, Phonons and Electrons}},\ in\ \href {https://doi.org/10.1007/978-3-540-74686-7_10} {\emph {\bibinfo {booktitle} {Computational Many-Particle Physics}}},\ \bibinfo {editor} {edited by\ \bibinfo {editor} {\bibfnamefont {H.}~\bibnamefont {Fehske}}, \bibinfo {editor} {\bibfnamefont {R.}~\bibnamefont {Schneider}},\ and\ \bibinfo {editor} {\bibfnamefont {A.}~\bibnamefont {Wei{\ss}e}}}\ (\bibinfo  {publisher} {Springer Berlin Heidelberg},\ \bibinfo {address} {Berlin, Heidelberg},\ \bibinfo {year} {2008})\ pp.\ \bibinfo {pages} {277--356}\BibitemShut {NoStop}%
\bibitem [{\citenamefont {Gubernatis}\ \emph {et~al.}(2016)\citenamefont {Gubernatis}, \citenamefont {Kawashima},\ and\ \citenamefont {Werner}}]{Gubernatis16}%
  \BibitemOpen
  \bibfield  {author} {\bibinfo {author} {\bibfnamefont {J.}~\bibnamefont {Gubernatis}}, \bibinfo {author} {\bibfnamefont {N.}~\bibnamefont {Kawashima}},\ and\ \bibinfo {author} {\bibfnamefont {P.}~\bibnamefont {Werner}},\ }\href {https://doi.org/10.1143/PTPS.145.138} {\emph {\bibinfo {title} {{Quantum Monte Carlo Methods: Algorithms for Lattice Models}}}}\ (\bibinfo  {publisher} {Cambridge University Press, Cambridge, England},\ \bibinfo {year} {2016})\BibitemShut {NoStop}%
\bibitem [{\citenamefont {Hirsch}(1983)}]{Hirsch1983}%
  \BibitemOpen
  \bibfield  {author} {\bibinfo {author} {\bibfnamefont {J.~E.}\ \bibnamefont {Hirsch}},\ }\bibfield  {title} {\bibinfo {title} {{Discrete Hubbard-Stratonovich transformation for fermion lattice models}},\ }\href {https://doi.org/10.1103/PhysRevB.28.4059} {\bibfield  {journal} {\bibinfo  {journal} {Phys. Rev. B}\ }\textbf {\bibinfo {volume} {28}},\ \bibinfo {pages} {4059} (\bibinfo {year} {1983})}\BibitemShut {NoStop}%
\bibitem [{\citenamefont {White}\ and\ \citenamefont {Wilkins}(1988)}]{White1988}%
  \BibitemOpen
  \bibfield  {author} {\bibinfo {author} {\bibfnamefont {S.~R.}\ \bibnamefont {White}}\ and\ \bibinfo {author} {\bibfnamefont {J.~W.}\ \bibnamefont {Wilkins}},\ }\bibfield  {title} {\bibinfo {title} {{Fermion simulations in systems with negative weights}},\ }\href {https://doi.org/10.1103/PhysRevB.37.5024} {\bibfield  {journal} {\bibinfo  {journal} {Phys. Rev. B}\ }\textbf {\bibinfo {volume} {37}},\ \bibinfo {pages} {5024} (\bibinfo {year} {1988})}\BibitemShut {NoStop}%
\bibitem [{\citenamefont {Loh}\ \emph {et~al.}(1990)\citenamefont {Loh}, \citenamefont {Gubernatis}, \citenamefont {Scalettar}, \citenamefont {White}, \citenamefont {Scalapino},\ and\ \citenamefont {Sugar}}]{Loh1990}%
  \BibitemOpen
  \bibfield  {author} {\bibinfo {author} {\bibfnamefont {E.~Y.}\ \bibnamefont {Loh}}, \bibinfo {author} {\bibfnamefont {J.~E.}\ \bibnamefont {Gubernatis}}, \bibinfo {author} {\bibfnamefont {R.~T.}\ \bibnamefont {Scalettar}}, \bibinfo {author} {\bibfnamefont {S.~R.}\ \bibnamefont {White}}, \bibinfo {author} {\bibfnamefont {D.~J.}\ \bibnamefont {Scalapino}},\ and\ \bibinfo {author} {\bibfnamefont {R.~L.}\ \bibnamefont {Sugar}},\ }\bibfield  {title} {\bibinfo {title} {{Sign problem in the numerical simulation of many-electron systems}},\ }\href {https://doi.org/10.1103/PhysRevB.41.9301} {\bibfield  {journal} {\bibinfo  {journal} {Phys. Rev. B}\ }\textbf {\bibinfo {volume} {41}},\ \bibinfo {pages} {9301} (\bibinfo {year} {1990})}\BibitemShut {NoStop}%
\bibitem [{\citenamefont {Mondaini}\ \emph {et~al.}(2022)\citenamefont {Mondaini}, \citenamefont {Tarat},\ and\ \citenamefont {Scalettar}}]{Mondaini2022}%
  \BibitemOpen
  \bibfield  {author} {\bibinfo {author} {\bibfnamefont {R.}~\bibnamefont {Mondaini}}, \bibinfo {author} {\bibfnamefont {S.}~\bibnamefont {Tarat}},\ and\ \bibinfo {author} {\bibfnamefont {R.~T.}\ \bibnamefont {Scalettar}},\ }\bibfield  {title} {\bibinfo {title} {{Quantum critical points and the sign problem}},\ }\href {https://doi.org/10.1126/science.abg9299} {\bibfield  {journal} {\bibinfo  {journal} {Science}\ }\textbf {\bibinfo {volume} {375}},\ \bibinfo {pages} {418} (\bibinfo {year} {2022})}\BibitemShut {NoStop}%
\bibitem [{\citenamefont {Mondaini}\ \emph {et~al.}(2023)\citenamefont {Mondaini}, \citenamefont {Tarat},\ and\ \citenamefont {Scalettar}}]{Mondaini2023}%
  \BibitemOpen
  \bibfield  {author} {\bibinfo {author} {\bibfnamefont {R.}~\bibnamefont {Mondaini}}, \bibinfo {author} {\bibfnamefont {S.}~\bibnamefont {Tarat}},\ and\ \bibinfo {author} {\bibfnamefont {R.~T.}\ \bibnamefont {Scalettar}},\ }\bibfield  {title} {\bibinfo {title} {{Universality and critical exponents of the fermion sign problem}},\ }\href {https://doi.org/10.1103/PhysRevB.107.245144} {\bibfield  {journal} {\bibinfo  {journal} {Phys. Rev. B}\ }\textbf {\bibinfo {volume} {107}},\ \bibinfo {pages} {245144} (\bibinfo {year} {2023})}\BibitemShut {NoStop}%
\bibitem [{\citenamefont {Kaul}(2015)}]{Kaul2015}%
  \BibitemOpen
  \bibfield  {author} {\bibinfo {author} {\bibfnamefont {R.~K.}\ \bibnamefont {Kaul}},\ }\bibfield  {title} {\bibinfo {title} {Spin nematics, valence-bond solids, and spin liquids in $\mathrm{SO}(n)$ quantum spin models on the triangular lattice},\ }\href {https://doi.org/10.1103/PhysRevLett.115.157202} {\bibfield  {journal} {\bibinfo  {journal} {Phys. Rev. Lett.}\ }\textbf {\bibinfo {volume} {115}},\ \bibinfo {pages} {157202} (\bibinfo {year} {2015})}\BibitemShut {NoStop}%
\bibitem [{\citenamefont {Sato}\ \emph {et~al.}(2018)\citenamefont {Sato}, \citenamefont {Assaad},\ and\ \citenamefont {Grover}}]{Sato2018}%
  \BibitemOpen
  \bibfield  {author} {\bibinfo {author} {\bibfnamefont {T.}~\bibnamefont {Sato}}, \bibinfo {author} {\bibfnamefont {F.~F.}\ \bibnamefont {Assaad}},\ and\ \bibinfo {author} {\bibfnamefont {T.}~\bibnamefont {Grover}},\ }\bibfield  {title} {\bibinfo {title} {{Quantum Monte Carlo Simulation of Frustrated Kondo Lattice Models}},\ }\href {https://doi.org/10.1103/PhysRevLett.120.107201} {\bibfield  {journal} {\bibinfo  {journal} {Phys. Rev. Lett.}\ }\textbf {\bibinfo {volume} {120}},\ \bibinfo {pages} {107201} (\bibinfo {year} {2018})}\BibitemShut {NoStop}%
\bibitem [{\citenamefont {Liu}\ \emph {et~al.}(2018)\citenamefont {Liu}, \citenamefont {Xu}, \citenamefont {Qi}, \citenamefont {Sun},\ and\ \citenamefont {Meng}}]{Liu2018}%
  \BibitemOpen
  \bibfield  {author} {\bibinfo {author} {\bibfnamefont {Z.~H.}\ \bibnamefont {Liu}}, \bibinfo {author} {\bibfnamefont {X.~Y.}\ \bibnamefont {Xu}}, \bibinfo {author} {\bibfnamefont {Y.}~\bibnamefont {Qi}}, \bibinfo {author} {\bibfnamefont {K.}~\bibnamefont {Sun}},\ and\ \bibinfo {author} {\bibfnamefont {Z.~Y.}\ \bibnamefont {Meng}},\ }\bibfield  {title} {\bibinfo {title} {Itinerant quantum critical point with frustration and a non-{F}ermi liquid},\ }\href {https://doi.org/10.1103/PhysRevB.98.045116} {\bibfield  {journal} {\bibinfo  {journal} {Physical Review B}\ }\textbf {\bibinfo {volume} {98}},\ \bibinfo {pages} {045116} (\bibinfo {year} {2018})}\BibitemShut {NoStop}%
\bibitem [{\citenamefont {\v{S}untajs}\ \emph {et~al.}(2020)\citenamefont {\v{S}untajs}, \citenamefont {Bon\v{c}a}, \citenamefont {Prosen},\ and\ \citenamefont {Vidmar}}]{Suntajs2020}%
  \BibitemOpen
  \bibfield  {author} {\bibinfo {author} {\bibfnamefont {J.}~\bibnamefont {\v{S}untajs}}, \bibinfo {author} {\bibfnamefont {J.}~\bibnamefont {Bon\v{c}a}}, \bibinfo {author} {\bibfnamefont {T.}~\bibnamefont {Prosen}},\ and\ \bibinfo {author} {\bibfnamefont {L.}~\bibnamefont {Vidmar}},\ }\bibfield  {title} {\bibinfo {title} {{Ergodicity breaking transition in finite disordered spin chains}},\ }\href {https://doi.org/10.1103/PhysRevB.102.064207} {\bibfield  {journal} {\bibinfo  {journal} {Phys. Rev. B}\ }\textbf {\bibinfo {volume} {102}},\ \bibinfo {pages} {064207} (\bibinfo {year} {2020})}\BibitemShut {NoStop}%
\bibitem [{\citenamefont {Jin}\ \emph {et~al.}(2022)\citenamefont {Jin}, \citenamefont {Liu}, \citenamefont {Mondaini},\ and\ \citenamefont {Rigol}}]{Jin2022}%
  \BibitemOpen
  \bibfield  {author} {\bibinfo {author} {\bibfnamefont {X.}~\bibnamefont {Jin}}, \bibinfo {author} {\bibfnamefont {Y.}~\bibnamefont {Liu}}, \bibinfo {author} {\bibfnamefont {R.}~\bibnamefont {Mondaini}},\ and\ \bibinfo {author} {\bibfnamefont {M.}~\bibnamefont {Rigol}},\ }\bibfield  {title} {\bibinfo {title} {Charge excitations across a superconductor-insulator transition},\ }\href {https://doi.org/10.1103/PhysRevB.106.245117} {\bibfield  {journal} {\bibinfo  {journal} {Phys. Rev. B}\ }\textbf {\bibinfo {volume} {106}},\ \bibinfo {pages} {245117} (\bibinfo {year} {2022})}\BibitemShut {NoStop}%
\bibitem [{\citenamefont {Sandvik}(2010)}]{Sandvik2010}%
  \BibitemOpen
  \bibfield  {author} {\bibinfo {author} {\bibfnamefont {A.~W.}\ \bibnamefont {Sandvik}},\ }\bibfield  {title} {\bibinfo {title} {Computational studies of quantum spin systems},\ }\href {https://doi.org/10.1063/1.3518900} {\bibfield  {journal} {\bibinfo  {journal} {AIP Conference Proceedings}\ }\textbf {\bibinfo {volume} {1297}},\ \bibinfo {pages} {135} (\bibinfo {year} {2010})}\BibitemShut {NoStop}%
\bibitem [{\citenamefont {Assaad}\ and\ \citenamefont {Herbut}(2013)}]{Assaad2013}%
  \BibitemOpen
  \bibfield  {author} {\bibinfo {author} {\bibfnamefont {F.~F.}\ \bibnamefont {Assaad}}\ and\ \bibinfo {author} {\bibfnamefont {I.~F.}\ \bibnamefont {Herbut}},\ }\bibfield  {title} {\bibinfo {title} {{Pinning the Order: The Nature of Quantum Criticality in the {H}ubbard Model on Honeycomb Lattice}},\ }\href {https://doi.org/10.1103/PhysRevX.3.031010} {\bibfield  {journal} {\bibinfo  {journal} {Phys. Rev. X}\ }\textbf {\bibinfo {volume} {3}},\ \bibinfo {pages} {031010} (\bibinfo {year} {2013})}\BibitemShut {NoStop}%
\bibitem [{\citenamefont {Parisen~Toldin}\ \emph {et~al.}(2015)\citenamefont {Parisen~Toldin}, \citenamefont {Hohenadler}, \citenamefont {Assaad},\ and\ \citenamefont {Herbut}}]{Parisen2015}%
  \BibitemOpen
  \bibfield  {author} {\bibinfo {author} {\bibfnamefont {F.}~\bibnamefont {Parisen~Toldin}}, \bibinfo {author} {\bibfnamefont {M.}~\bibnamefont {Hohenadler}}, \bibinfo {author} {\bibfnamefont {F.~F.}\ \bibnamefont {Assaad}},\ and\ \bibinfo {author} {\bibfnamefont {I.~F.}\ \bibnamefont {Herbut}},\ }\bibfield  {title} {\bibinfo {title} {Fermionic quantum criticality in honeycomb and $\ensuremath{\pi}$-flux {H}ubbard models: Finite-size scaling of renormalization-group-invariant observables from quantum {M}onte {C}arlo},\ }\href {https://doi.org/10.1103/PhysRevB.91.165108} {\bibfield  {journal} {\bibinfo  {journal} {Phys. Rev. B}\ }\textbf {\bibinfo {volume} {91}},\ \bibinfo {pages} {165108} (\bibinfo {year} {2015})}\BibitemShut {NoStop}%
\bibitem [{\citenamefont {Otsuka}\ \emph {et~al.}(2016)\citenamefont {Otsuka}, \citenamefont {Yunoki},\ and\ \citenamefont {Sorella}}]{Otsuka2016}%
  \BibitemOpen
  \bibfield  {author} {\bibinfo {author} {\bibfnamefont {Y.}~\bibnamefont {Otsuka}}, \bibinfo {author} {\bibfnamefont {S.}~\bibnamefont {Yunoki}},\ and\ \bibinfo {author} {\bibfnamefont {S.}~\bibnamefont {Sorella}},\ }\bibfield  {title} {\bibinfo {title} {{Universal Quantum Criticality in the Metal-Insulator Transition of Two-Dimensional Interacting {D}irac Electrons}},\ }\href {https://doi.org/10.1103/PhysRevX.6.011029} {\bibfield  {journal} {\bibinfo  {journal} {Phys. Rev. X}\ }\textbf {\bibinfo {volume} {6}},\ \bibinfo {pages} {011029} (\bibinfo {year} {2016})}\BibitemShut {NoStop}%
\bibitem [{\citenamefont {Tang}\ \emph {et~al.}(2018)\citenamefont {Tang}, \citenamefont {Leaw}, \citenamefont {Rodrigues}, \citenamefont {Herbut}, \citenamefont {Sengupta}, \citenamefont {Assaad},\ and\ \citenamefont {Adam}}]{Tang2018}%
  \BibitemOpen
  \bibfield  {author} {\bibinfo {author} {\bibfnamefont {H.-K.}\ \bibnamefont {Tang}}, \bibinfo {author} {\bibfnamefont {J.}~\bibnamefont {Leaw}}, \bibinfo {author} {\bibfnamefont {J.}~\bibnamefont {Rodrigues}}, \bibinfo {author} {\bibfnamefont {I.}~\bibnamefont {Herbut}}, \bibinfo {author} {\bibfnamefont {P.}~\bibnamefont {Sengupta}}, \bibinfo {author} {\bibfnamefont {F.}~\bibnamefont {Assaad}},\ and\ \bibinfo {author} {\bibfnamefont {S.}~\bibnamefont {Adam}},\ }\bibfield  {title} {\bibinfo {title} {{The role of electron-electron interactions in two-dimensional Dirac fermions}},\ }\href {https://doi.org/10.1126/science.aao2934} {\bibfield  {journal} {\bibinfo  {journal} {Science}\ }\textbf {\bibinfo {volume} {361}},\ \bibinfo {pages} {570} (\bibinfo {year} {2018})}\BibitemShut {NoStop}%
\bibitem [{\citenamefont {Otsuka}\ \emph {et~al.}(2020)\citenamefont {Otsuka}, \citenamefont {Seki}, \citenamefont {Sorella},\ and\ \citenamefont {Yunoki}}]{Otsuka2020}%
  \BibitemOpen
  \bibfield  {author} {\bibinfo {author} {\bibfnamefont {Y.}~\bibnamefont {Otsuka}}, \bibinfo {author} {\bibfnamefont {K.}~\bibnamefont {Seki}}, \bibinfo {author} {\bibfnamefont {S.}~\bibnamefont {Sorella}},\ and\ \bibinfo {author} {\bibfnamefont {S.}~\bibnamefont {Yunoki}},\ }\bibfield  {title} {\bibinfo {title} {Dirac electrons in the square-lattice hubbard model with a $d$-wave pairing field: The chiral {H}eisenberg universality class revisited},\ }\href {https://doi.org/10.1103/PhysRevB.102.235105} {\bibfield  {journal} {\bibinfo  {journal} {Phys. Rev. B}\ }\textbf {\bibinfo {volume} {102}},\ \bibinfo {pages} {235105} (\bibinfo {year} {2020})}\BibitemShut {NoStop}%
\bibitem [{\citenamefont {Kennedy}\ \emph {et~al.}(2025)\citenamefont {Kennedy}, \citenamefont {Sousa-J\'unior}, \citenamefont {Costa},\ and\ \citenamefont {dos Santos}}]{Kennedy2025}%
  \BibitemOpen
  \bibfield  {author} {\bibinfo {author} {\bibfnamefont {W.}~\bibnamefont {Kennedy}}, \bibinfo {author} {\bibfnamefont {S.~a. d.~A.}\ \bibnamefont {Sousa-J\'unior}}, \bibinfo {author} {\bibfnamefont {N.~C.}\ \bibnamefont {Costa}},\ and\ \bibinfo {author} {\bibfnamefont {R.~R.}\ \bibnamefont {dos Santos}},\ }\bibfield  {title} {\bibinfo {title} {Extended {H}ubbard model on a honeycomb lattice},\ }\href {https://doi.org/10.1103/PhysRevB.111.155118} {\bibfield  {journal} {\bibinfo  {journal} {Phys. Rev. B}\ }\textbf {\bibinfo {volume} {111}},\ \bibinfo {pages} {155118} (\bibinfo {year} {2025})}\BibitemShut {NoStop}%
\bibitem [{\citenamefont {Assaad}\ \emph {et~al.}(2022)\citenamefont {Assaad}, \citenamefont {Bercx}, \citenamefont {Goth}, \citenamefont {Götz}, \citenamefont {Hofmann}, \citenamefont {Huffman}, \citenamefont {Liu}, \citenamefont {Toldin}, \citenamefont {Portela},\ and\ \citenamefont {Schwab}}]{Assaad2022}%
  \BibitemOpen
  \bibfield  {author} {\bibinfo {author} {\bibfnamefont {F.~F.}\ \bibnamefont {Assaad}}, \bibinfo {author} {\bibfnamefont {M.}~\bibnamefont {Bercx}}, \bibinfo {author} {\bibfnamefont {F.}~\bibnamefont {Goth}}, \bibinfo {author} {\bibfnamefont {A.}~\bibnamefont {Götz}}, \bibinfo {author} {\bibfnamefont {J.~S.}\ \bibnamefont {Hofmann}}, \bibinfo {author} {\bibfnamefont {E.}~\bibnamefont {Huffman}}, \bibinfo {author} {\bibfnamefont {Z.}~\bibnamefont {Liu}}, \bibinfo {author} {\bibfnamefont {F.~P.}\ \bibnamefont {Toldin}}, \bibinfo {author} {\bibfnamefont {J.~S.~E.}\ \bibnamefont {Portela}},\ and\ \bibinfo {author} {\bibfnamefont {J.}~\bibnamefont {Schwab}},\ }\bibfield  {title} {\bibinfo {title} {{The ALF (Algorithms for Lattice Fermions) project release 2.0. Documentation for the auxiliary-field quantum Monte Carlo code}},\ }\href {https://doi.org/10.21468/SciPostPhysCodeb.1} {\bibfield  {journal} {\bibinfo  {journal} {SciPost Phys. Codebases}\ ,\ \bibinfo {pages} {1}} (\bibinfo {year} {2022})}\BibitemShut
  {NoStop}%
\bibitem [{\citenamefont {Yin}\ \emph {et~al.}(2022)\citenamefont {Yin}, \citenamefont {Du}, \citenamefont {Zhang}, \citenamefont {Chen}, \citenamefont {Pei}, \citenamefont {Zhou}, \citenamefont {Gu}, \citenamefont {Xu}, \citenamefont {Zhang}, \citenamefont {Zhao}, \citenamefont {Li}, \citenamefont {Xu}, \citenamefont {Bernevig}, \citenamefont {Liu}, \citenamefont {Liu}, \citenamefont {Chen},\ and\ \citenamefont {Yang}}]{Yin2022}%
  \BibitemOpen
  \bibfield  {author} {\bibinfo {author} {\bibfnamefont {Z.~X.}\ \bibnamefont {Yin}}, \bibinfo {author} {\bibfnamefont {X.}~\bibnamefont {Du}}, \bibinfo {author} {\bibfnamefont {S.}~\bibnamefont {Zhang}}, \bibinfo {author} {\bibfnamefont {C.}~\bibnamefont {Chen}}, \bibinfo {author} {\bibfnamefont {D.}~\bibnamefont {Pei}}, \bibinfo {author} {\bibfnamefont {J.~S.}\ \bibnamefont {Zhou}}, \bibinfo {author} {\bibfnamefont {X.}~\bibnamefont {Gu}}, \bibinfo {author} {\bibfnamefont {R.~Z.}\ \bibnamefont {Xu}}, \bibinfo {author} {\bibfnamefont {Q.~Q.}\ \bibnamefont {Zhang}}, \bibinfo {author} {\bibfnamefont {W.~X.}\ \bibnamefont {Zhao}}, \bibinfo {author} {\bibfnamefont {Y.~D.}\ \bibnamefont {Li}}, \bibinfo {author} {\bibfnamefont {Y.~F.}\ \bibnamefont {Xu}}, \bibinfo {author} {\bibfnamefont {A.}~\bibnamefont {Bernevig}}, \bibinfo {author} {\bibfnamefont {Z.~K.}\ \bibnamefont {Liu}}, \bibinfo {author} {\bibfnamefont {E.~K.}\ \bibnamefont {Liu}}, \bibinfo {author} {\bibfnamefont {Y.~L.}\ \bibnamefont {Chen}},\ and\
  \bibinfo {author} {\bibfnamefont {L.~X.}\ \bibnamefont {Yang}},\ }\bibfield  {title} {\bibinfo {title} {Electronic structure of antiferromagnetic {D}irac semimetal candidate {GdIn}$_3$},\ }\href {https://doi.org/10.1103/PhysRevMaterials.6.084203} {\bibfield  {journal} {\bibinfo  {journal} {Phys. Rev. Mater.}\ }\textbf {\bibinfo {volume} {6}},\ \bibinfo {pages} {084203} (\bibinfo {year} {2022})}\BibitemShut {NoStop}%
\bibitem [{\citenamefont {Masuda}\ \emph {et~al.}(2016)\citenamefont {Masuda}, \citenamefont {Sakai}, \citenamefont {Tokunaga}, \citenamefont {Yamasaki}, \citenamefont {Miyake}, \citenamefont {Shiogai}, \citenamefont {Nakamura}, \citenamefont {Awaji}, \citenamefont {Tsukazaki}, \citenamefont {Nakao}, \citenamefont {Murakami}, \citenamefont {hisa Arima}, \citenamefont {Tokura},\ and\ \citenamefont {Ishiwata}}]{Masuda2016}%
  \BibitemOpen
  \bibfield  {author} {\bibinfo {author} {\bibfnamefont {H.}~\bibnamefont {Masuda}}, \bibinfo {author} {\bibfnamefont {H.}~\bibnamefont {Sakai}}, \bibinfo {author} {\bibfnamefont {M.}~\bibnamefont {Tokunaga}}, \bibinfo {author} {\bibfnamefont {Y.}~\bibnamefont {Yamasaki}}, \bibinfo {author} {\bibfnamefont {A.}~\bibnamefont {Miyake}}, \bibinfo {author} {\bibfnamefont {J.}~\bibnamefont {Shiogai}}, \bibinfo {author} {\bibfnamefont {S.}~\bibnamefont {Nakamura}}, \bibinfo {author} {\bibfnamefont {S.}~\bibnamefont {Awaji}}, \bibinfo {author} {\bibfnamefont {A.}~\bibnamefont {Tsukazaki}}, \bibinfo {author} {\bibfnamefont {H.}~\bibnamefont {Nakao}}, \bibinfo {author} {\bibfnamefont {Y.}~\bibnamefont {Murakami}}, \bibinfo {author} {\bibfnamefont {T.}~\bibnamefont {hisa Arima}}, \bibinfo {author} {\bibfnamefont {Y.}~\bibnamefont {Tokura}},\ and\ \bibinfo {author} {\bibfnamefont {S.}~\bibnamefont {Ishiwata}},\ }\bibfield  {title} {\bibinfo {title} {Quantum {H}all effect in a bulk antiferromagnet {EuMnBi}$_2$ with
  magnetically confined two-dimensional {D}irac fermions},\ }\href {https://doi.org/10.1126/sciadv.1501117} {\bibfield  {journal} {\bibinfo  {journal} {Science Advances}\ }\textbf {\bibinfo {volume} {2}},\ \bibinfo {pages} {e1501117} (\bibinfo {year} {2016})}\BibitemShut {NoStop}%
\bibitem [{\citenamefont {Li}\ \emph {et~al.}(2019)\citenamefont {Li}, \citenamefont {Li}, \citenamefont {Du}, \citenamefont {Wang}, \citenamefont {Gu}, \citenamefont {Zhang}, \citenamefont {He}, \citenamefont {Duan},\ and\ \citenamefont {Xu}}]{Li2019}%
  \BibitemOpen
  \bibfield  {author} {\bibinfo {author} {\bibfnamefont {J.}~\bibnamefont {Li}}, \bibinfo {author} {\bibfnamefont {Y.}~\bibnamefont {Li}}, \bibinfo {author} {\bibfnamefont {S.}~\bibnamefont {Du}}, \bibinfo {author} {\bibfnamefont {Z.}~\bibnamefont {Wang}}, \bibinfo {author} {\bibfnamefont {B.-L.}\ \bibnamefont {Gu}}, \bibinfo {author} {\bibfnamefont {S.-C.}\ \bibnamefont {Zhang}}, \bibinfo {author} {\bibfnamefont {K.}~\bibnamefont {He}}, \bibinfo {author} {\bibfnamefont {W.}~\bibnamefont {Duan}},\ and\ \bibinfo {author} {\bibfnamefont {Y.}~\bibnamefont {Xu}},\ }\bibfield  {title} {\bibinfo {title} {Intrinsic magnetic topological insulators in van der {W}aals layered {MnBi$_2$Te$_4$}-family materials},\ }\href {https://doi.org/10.1126/sciadv.aaw5685} {\bibfield  {journal} {\bibinfo  {journal} {Science Advances}\ }\textbf {\bibinfo {volume} {5}},\ \bibinfo {pages} {eaaw5685} (\bibinfo {year} {2019})}\BibitemShut {NoStop}%
\bibitem [{\citenamefont {Li}\ \emph {et~al.}(2020)\citenamefont {Li}, \citenamefont {Yan}, \citenamefont {Pajerowski}, \citenamefont {Gordon}, \citenamefont {Nedi\ifmmode~\acute{c}\else \'{c}\fi{}}, \citenamefont {Sizyuk}, \citenamefont {Ke}, \citenamefont {Orth}, \citenamefont {Vaknin},\ and\ \citenamefont {McQueeney}}]{Li2020}%
  \BibitemOpen
  \bibfield  {author} {\bibinfo {author} {\bibfnamefont {B.}~\bibnamefont {Li}}, \bibinfo {author} {\bibfnamefont {J.-Q.}\ \bibnamefont {Yan}}, \bibinfo {author} {\bibfnamefont {D.~M.}\ \bibnamefont {Pajerowski}}, \bibinfo {author} {\bibfnamefont {E.}~\bibnamefont {Gordon}}, \bibinfo {author} {\bibfnamefont {A.-M.}\ \bibnamefont {Nedi\ifmmode~\acute{c}\else \'{c}\fi{}}}, \bibinfo {author} {\bibfnamefont {Y.}~\bibnamefont {Sizyuk}}, \bibinfo {author} {\bibfnamefont {L.}~\bibnamefont {Ke}}, \bibinfo {author} {\bibfnamefont {P.~P.}\ \bibnamefont {Orth}}, \bibinfo {author} {\bibfnamefont {D.}~\bibnamefont {Vaknin}},\ and\ \bibinfo {author} {\bibfnamefont {R.~J.}\ \bibnamefont {McQueeney}},\ }\bibfield  {title} {\bibinfo {title} {Competing magnetic interactions in the antiferromagnetic topological insulator {MnBi}$_2${Te}$_4$},\ }\href {https://doi.org/10.1103/PhysRevLett.124.167204} {\bibfield  {journal} {\bibinfo  {journal} {Phys. Rev. Lett.}\ }\textbf {\bibinfo {volume} {124}},\ \bibinfo {pages} {167204}
  (\bibinfo {year} {2020})}\BibitemShut {NoStop}%
\bibitem [{\citenamefont {Yang}\ \emph {et~al.}(2020)\citenamefont {Yang}, \citenamefont {Zhong},\ and\ \citenamefont {Luo}}]{Yang2020}%
  \BibitemOpen
  \bibfield  {author} {\bibinfo {author} {\bibfnamefont {W.-W.}\ \bibnamefont {Yang}}, \bibinfo {author} {\bibfnamefont {Y.}~\bibnamefont {Zhong}},\ and\ \bibinfo {author} {\bibfnamefont {H.-G.}\ \bibnamefont {Luo}},\ }\bibfield  {title} {\bibinfo {title} {Hexagonal {Ising-Kondo} lattice: {A}n implication for intrinsic antiferromagnetic topological insulator},\ }\href {https://doi.org/10.1103/PhysRevB.102.195141} {\bibfield  {journal} {\bibinfo  {journal} {Phys. Rev. B}\ }\textbf {\bibinfo {volume} {102}},\ \bibinfo {pages} {195141} (\bibinfo {year} {2020})}\BibitemShut {NoStop}%
\bibitem [{\citenamefont {Klett}\ \emph {et~al.}(2020)\citenamefont {Klett}, \citenamefont {Ok}, \citenamefont {Riegler}, \citenamefont {W\"olfle}, \citenamefont {Thomale},\ and\ \citenamefont {Neupert}}]{Klett2020}%
  \BibitemOpen
  \bibfield  {author} {\bibinfo {author} {\bibfnamefont {M.}~\bibnamefont {Klett}}, \bibinfo {author} {\bibfnamefont {S.}~\bibnamefont {Ok}}, \bibinfo {author} {\bibfnamefont {D.}~\bibnamefont {Riegler}}, \bibinfo {author} {\bibfnamefont {P.}~\bibnamefont {W\"olfle}}, \bibinfo {author} {\bibfnamefont {R.}~\bibnamefont {Thomale}},\ and\ \bibinfo {author} {\bibfnamefont {T.}~\bibnamefont {Neupert}},\ }\bibfield  {title} {\bibinfo {title} {Topology and magnetism in the {K}ondo insulator phase diagram},\ }\href {https://doi.org/10.1103/PhysRevB.101.161112} {\bibfield  {journal} {\bibinfo  {journal} {Phys. Rev. B}\ }\textbf {\bibinfo {volume} {101}},\ \bibinfo {pages} {161112} (\bibinfo {year} {2020})}\BibitemShut {NoStop}%
\bibitem [{\citenamefont {Ido}\ and\ \citenamefont {Misawa}(2024)}]{Ido2024}%
  \BibitemOpen
  \bibfield  {author} {\bibinfo {author} {\bibfnamefont {K.}~\bibnamefont {Ido}}\ and\ \bibinfo {author} {\bibfnamefont {T.}~\bibnamefont {Misawa}},\ }\bibfield  {title} {\bibinfo {title} {Many-body {C}hern insulator in the {K}ondo lattice model on a triangular lattice},\ }\href {https://doi.org/10.1103/PhysRevB.109.245114} {\bibfield  {journal} {\bibinfo  {journal} {Phys. Rev. B}\ }\textbf {\bibinfo {volume} {109}},\ \bibinfo {pages} {245114} (\bibinfo {year} {2024})}\BibitemShut {NoStop}%
\bibitem [{\citenamefont {dos Anjos Sousa-Júnior}(2025)}]{zenodo}%
  \BibitemOpen
  \bibfield  {author} {\bibinfo {author} {\bibfnamefont {S.}~\bibnamefont {dos Anjos Sousa-Júnior}},\ }\bibfield  {title} {\bibinfo {title} {Data from spin-orbit coupled periodic anderson model: Kondo-dirac semimetal and orbital-selective antiferromagnetic semimetal},\ }\href {https://doi.org/10.5281/zenodo.17613334} {10.5281/zenodo.17613334} (\bibinfo {year} {2025})\BibitemShut {NoStop}%
\bibitem [{\citenamefont {Zheng}\ \emph {et~al.}(2011)\citenamefont {Zheng}, \citenamefont {Zhang},\ and\ \citenamefont {Wu}}]{Zheng2011}%
  \BibitemOpen
  \bibfield  {author} {\bibinfo {author} {\bibfnamefont {D.}~\bibnamefont {Zheng}}, \bibinfo {author} {\bibfnamefont {G.-M.}\ \bibnamefont {Zhang}},\ and\ \bibinfo {author} {\bibfnamefont {C.}~\bibnamefont {Wu}},\ }\bibfield  {title} {\bibinfo {title} {{Particle-hole symmetry and interaction effects in the Kane-Mele-Hubbard model}},\ }\href {https://doi.org/10.1103/PhysRevB.84.205121} {\bibfield  {journal} {\bibinfo  {journal} {Phys. Rev. B}\ }\textbf {\bibinfo {volume} {84}},\ \bibinfo {pages} {205121} (\bibinfo {year} {2011})}\BibitemShut {NoStop}%
\bibitem [{\citenamefont {Hu}\ \emph {et~al.}(2017)\citenamefont {Hu}, \citenamefont {Scalettar}, \citenamefont {Huang},\ and\ \citenamefont {Moritz}}]{Hu2017}%
  \BibitemOpen
  \bibfield  {author} {\bibinfo {author} {\bibfnamefont {W.}~\bibnamefont {Hu}}, \bibinfo {author} {\bibfnamefont {R.~T.}\ \bibnamefont {Scalettar}}, \bibinfo {author} {\bibfnamefont {E.~W.}\ \bibnamefont {Huang}},\ and\ \bibinfo {author} {\bibfnamefont {B.}~\bibnamefont {Moritz}},\ }\bibfield  {title} {\bibinfo {title} {{Effects of an additional conduction band on the singlet-antiferromagnet competition in the periodic Anderson model}},\ }\href {https://doi.org/10.1103/PhysRevB.95.235122} {\bibfield  {journal} {\bibinfo  {journal} {Phys. Rev. B}\ }\textbf {\bibinfo {volume} {95}},\ \bibinfo {pages} {235122} (\bibinfo {year} {2017})}\BibitemShut {NoStop}%
\end{thebibliography}%

\end{document}